\newcommand{\be}{\begin{equation}}
\newcommand{\ee}{\end{equation}}

\documentstyle[psfig,12pt]{article}
\def\comma{ \; , }
\def\period{ \; . }
\def\Del{ \nabla }
\def\di{ \partial }
\def\dir{ \partial_r }
\def\minus{ \mbox - \, }
\def\half{ \frac{1}{2} }
\def\quarter{ \frac{1}{4} }
\def\RdBk#1{ \left(\,#1\,\right) }
\def\SqBk#1{ \left[\,#1\,\right] }
\def\BrBk#1{ \left\{\,#1\,\right\} }
\def\SOL#1{ {\cal L}\!\left[\,#1\,\right] }

\def\W#1#2{ W\!\left(\,#1,#2\,\right) }

\begin{document}
\onecolumn
\begin{flushright}
  WATPHYS TH-96/20 \\
  gr-qc/9612026
\end{flushright}
\vfill
\begin{center}
  {\Large \bf Scalar Wave Falloff in Asymptotically 
    Anti-de Sitter Backgrounds}
  \\ \vfill

  J. S. F. Chan$^{(1)}$ and R. B. Mann$^{(1,2)}$ \\
  \vspace{2cm}
  (1) Department of Applied Mathematics,
  University of Waterloo, Waterloo, Ontario, Canada N2L 3G1 \\
  (2) Department of Physics,
  University of Waterloo, Waterloo, Ontario, Canada N2L 3G1 \\
  \vspace{2cm}
  \vfill

  \begin{abstract}
    Conformally invariant scalar waves in black hole spacetimes
    which are asymptotically anti-de Sitter are investigated. We
    consider both the $(2+1)$-dimensional black hole and
    $(3+1)$-dimensional Schwarzschild-anti-de Sitter spacetime as
    backgrounds. Analytical and numerical methods show that the
    waves decay exponentially in the $(2+1)$ dimensional black
    hole background. However the falloff pattern of the conformal
    scalar waves in the Schwarzschild-anti-de Sitter background is
    generally neither exponential nor an inverse power rate, although
    the approximate falloff of the maximal peak is weakly exponential. 
    We discuss the  implications of these results for mass inflation.
  \end{abstract}
  \vfill
\end{center}
\clearpage

\section{Introduction}

  It is well known that the maximally extended Reissner-Nordstr\"om
  spacetime can be imagined as a collection of different
  asymptotically flat universes connected by different charged
  black holes \cite{Hawking}. Except for the Schwarzschild
  solution, all the special solutions of the more general
  Kerr-Newman class of spacetimes can have two horizons, the inner
  and outer horizons. Nevertheless gravitational theorists find
  that these dual-horizon black holes are unphysical because
  causality can be violated inside the hole \cite{Carter}.
  Moreover any radiation (either electromagnetic or gravitational
  in nature) that goes into this kind of black hole will be
  indefinitely blue-shifted at the inner (or Cauchy) horizon
  \cite{Hawking}. This effect has caused some to expect that this
  null hypersurface acts like a barricade to the other universes
  in the maximally extended spacetime.

  This infinite blue-shift phenomenon at the Cauchy horizon was
  first discussed by Penrose in the late sixties \cite{Penrose}.
  At that time, people believed that any small energy perturbation
  on these dual-horizon black holes would destroy the Cauchy
  horizon because the perturbation is indefinitely magnified
  there, causing an infinite spacetime curvature at the horizon.
  Thus the null hypersurface would become a spacelike curvature
  singularity and the gateway to other universes is sealed. Both
  numerical and analytic approaches \cite{Simpson,Chandrasekhar}
  suggest that this null hypersurface is perturbatively unstable.
  However, although singularities are found at the Cauchy horizon,
  they are not spacetime curvature singularities at all. By taking
  the diverging stress-energy tensor into account, Hiscock showed
  that the perturbation only turns the horizon into a so-called
  {\em whimper singularity}: all curvature scalars are finite but
  a freely falling observer crossing the horizon measures an
  infinite energy density \cite{Hiscock}. This kind of singularity
  is too mild to seal off the passage to other universes, and so
  resolution of the issue would necessitate a study that did not
  rely on perturbation theory.

  Poisson and Israel made a breakthrough in this problem by
  showing that the Cauchy horizon can turn into a scalar spacetime
  curvature singularity \cite{Poisson}. Unlike Hiscock who
  considered a Reissner-Nordstr\"om black hole irradiated by a
  flux of incoming radiation, Poisson and Israel imposed both
  incoming and outgoing fluxes of radiation on a Reissner-Nordstr\"om
  background. The outgoing flux, even if negligibly small in
  quantity, makes the inner mass function of the black hole
  inflate without bound at the Cauchy horizon. More precisely, the
  inner mass of the black hole diverges at a rate of
  $\exp( \kappa\,v )/v^p$ near the Cauchy horizon. The factor
  $1/v^p$ comes from the decay rate of the scattered radiation
  tail \cite{Poisson,Ori1,Price}, where $p > 0$ and
  $v \rightarrow \infty$ at the Cauchy horizon. Regardless of the
  values of $p$ and the surface gravity $\kappa > 0$, the mass
  parameter always grows although the exponential rate is
  attenuated by the decaying effect of the radiation tail. This
  phenomenon is called {\em mass inflation} and expected to seal
  the inner horizon because the diverging mass parameter induces a
  scalar curvature singularity at the horizon. This result implies
  that it is inappropriate to maximally extend any dual-horizon
  black holes beyond the Cauchy horizon because the spacetime is
  unstable against energy perturbations there. It is generally
  believed that such perturbations (in the form of gravitational
  radiation) always exist in more realistic black holes which does
  not have perfect spherical or axial symmetry. These are
  scattered around the black hole forming incoming and outgoing
  fluxes but they will eventually decay away as a tail of late
  time radiation \cite{Price}. In this way, mass inflation is
  expected to prevent violation of causality.

  In addition to the Reissner-Nordstr\"om black hole, the mass
  inflation phenomenon has been found to take place in other black
  hole configurations \cite{Ori2,Droz,MI1,Husain,Cai,Bonanno,MI2}.
  These configurations are in $1+1$, $2+1$ and $3+1$ dimensions as
  well as in asymptotically non-flat spacetimes. All of these
  calculations assumed the inverse power-law decay for late time
  radiation as an ansatz to obtain the inflating mass function
  near the Cauchy horizon. Since the mass parameter is attenuated
  by the decaying effect of the radiation tail, it is important to
  understand the behaviour of the radiative tail in a spacetime
  other than Reissner-Nordstr\"om class. Mellor and Moss have
  shown that the radiation from perturbations in a de Sitter
  background exponentially decreases \cite{Mellor}. Strictly
  speaking this result has nothing to do with late time falloff
  because the global geometry extends beyond the cosmological
  horizon. However it indicates that the radiative falloff
  behaviour is sensitive to the presence of the cosmological
  constant. More recent work by Ching {\sl et al.} \cite{Ching1}
  demonstrated that under certain circumstances the tail can be
  something other than the simple inverse power-law.

  In this paper we will study radiative falloff in spacetimes that
  are not asymptotically flat. We find that the inverse power-law
  \cite{Price} is not universally true, and that in some
  asymptotically anti-de Sitter spacetimes, the late time tail
  decays exponentially. The asymptotically anti-de Sitter
  backgrounds we will study are the $(2+1)$-dimensional black hole
  \cite{BTZ} and Schwarzschild-anti-de Sitter spacetime. 

  The outline of our paper is as follows. In section 2 we review
  the structure of the $(D+1)$-dimensional scalar wave equation in
  spherically symmetric spacetimes that are not necessarily
  asymptotically flat, and discuss our numerical approach towards
  solving it. In section 3 we verify that our numerical approach
  correctly reproduces the power-law falloff in asymptotically
  flat spacetimes, and we cross-check this analytically. In the
  next two sections we study the falloff behaviour in the
  $(2+1)$-dimensional (or 3D) black hole background and in
  $(3+1)$-dimensional Schwarzschild-anti-de Sitter spacetime.
  Concluding remarks and an appendix round out our work.

\section{The Wave Equation in $D+1$ Dimensions}
  We shall study scalar waves in different dimensions, since wave
  equations for higher-spin fields are of a qualitatively similar
  structure \cite{Zerilli}. The (conformally coupled) scalar wave
  equation in $D+1$ dimensions is
  \begin{eqnarray}
    \Del^2 \Psi & = & \xi\,R\,\Psi \comma \label{WE1}
  \end{eqnarray}
  where $\xi$ is an arbitrary constant. If $\xi = (D-1)/(4D)$,
  this equation is conformally invariant. We simplify the problem
  by considering only static, spherically symmetric
  $(D+1)$-dimensional spacetimes with metric 
  \begin{eqnarray}
    ds^2 & = &
    \minus N(r)\,dt^2 + \frac{dr^2}{N(r)} + r^2\,d{\Omega_{D-1}}^2
    \comma \label{MetricForm}
  \end{eqnarray}
  where $N(r)$ is the lapse function and $d{\Omega_{D-1}}^2$ is
  the metric of a $(D-1)$-dimensional unit sphere. Generalizing
  Zerilli's separability technique \cite{Zerilli} to $D$
  dimensions, we assume
  \begin{eqnarray}
    \Psi & = & r^{(1-D)/2}\,\psi(t,r)\,Y^D_l \period
  \end{eqnarray}
  The functions $Y^D_l$ are the $D$-dimensional spherical
  harmonics which satisfy the equation
  \begin{eqnarray}
    \hat{L}^2\!\SqBk{Y^D_l} & = & \minus l\,(l+D-2)\,Y^D_l \period
  \end{eqnarray}
  The product $\minus l\,(l+D-2)$ is the eigenvalue of the
  operator $\hat{L}^2$ which is the angular derivative operator.
  It is straightforward to show that equation (\ref{WE1}) gives
  \begin{eqnarray}
    \minus \di_{tt} \psi(t,r) + N(r)\,\dir \SqBk{N(r)\,\dir \psi(t,r)}
    - N(r)\,V_e(r)\,\psi(t,r)
    & = &
    0 \period \label{WE2}
  \end{eqnarray}
  The function $V_e(r)$ is defined as
  \begin{eqnarray}
    V_e(r) & \equiv &
    \xi\,R + \frac{D-1}{2\,r}\,\frac{d}{d\,r} N(r)
    + \frac{(D-1)\,(D-3)}{4\,r^2}\,\SqBk{N(r) - 1}
    + \frac{(2\,l+D-3)\,(2\,l+D-1)}{4\,r^2} \period \label{Ve}
  \end{eqnarray}
  One can rewrite the wave equation (\ref{WE2}) as
  \begin{eqnarray}
    \minus \di_{tt} \psi(t,r) + N(r)\,\SOL{\psi(t,r)} & = & 0
    \comma \label{WE3}
  \end{eqnarray}
  with the help of a spatial differential operator
  \begin{eqnarray}
    {\cal L} & \equiv & \dir \SqBk{N(r)\,\dir} - V_e(r) \period
  \end{eqnarray}
  Alternatively, if we introduce
  \begin{equation}
    x \; \equiv \; \int \frac{dr}{N(r)} \comma
  \end{equation}
  then (\ref{WE2}) can be written as
  \begin{eqnarray}
    \di_{tt} \psi\!\RdBk{t,r(x)} - \di_{xx} \psi\!\RdBk{t,r(x)}
    + V\!\RdBk{r(x)}\,\psi\!\RdBk{t,r(x)} & = & 0 \period \label{WE5}
  \end{eqnarray}
  or as
  \begin{eqnarray}
    \di_{uv} \psi(u,v) & = &
    \minus \quarter\,N(r(u,v))\,V_e(r(u,v))\,\psi(u,v) \label{WE4}
  \end{eqnarray}
  using null co-ordinates $u = t - x$ and $v=t+x$. The function
  $V(r)$ (defined as $V(r) \equiv N(r)\,V_e(r)$) plays the role of
  a potential barrier which is induced from the background
  spacetime geometry. Although the potential when written in terms
  of the tortoise co-ordinate $x$ can be very complicated,
  equation (\ref{WE5}) has the familiar form of a potential
  scattering problem.

  Equation (\ref{WE5}) can be integrated numerically in a
  straightforward fashion by using finite difference method. First
  of all the D'Alembert operator $\di_{tt} - \di_{xx}$ can be
  discretized as
  \begin{eqnarray}
    \frac{\psi(t-\Delta t,x)-2\,\psi(t,x)+\psi(t+\Delta t,x)}{{\Delta t}^2}
    - \frac{\psi(t,x-\Delta x)-2\,\psi(t,x)+\psi(t,x+\Delta x)}{{\Delta x}^2}
    + O({\Delta t}^2) + O({\Delta x}^2)
  \end{eqnarray}
  using Taylor's theorem. In order to formulate a well-posed
  Cauchy problem we need to include the initial conditions, which
  for simplicity we choose to be
  \begin{eqnarray}\label{inicond}
    \psi(t=0,x) \; = \; 0
    \hspace{1cm} & {\rm and} & \hspace{1cm}
    \di_t \psi(t=0,x) \; = \; u(x) \period
  \end{eqnarray}
  Because the field $\psi$ is initially zero, its subsequent
  evolution is solely the result of the initial impulse of the
  field $\di_t \psi$. Discretizing the second condition in
  (\ref{inicond}) yields
  \begin{eqnarray}
    \frac{\psi(\Delta t,x)-\psi(\minus \Delta t,x)}{2\,\Delta t}
    & = & u(x) + O({\Delta t}^2) \comma
  \end{eqnarray}
  where we employ a Gaussian distribution with finite support for
  $u(x)$. We further define
  \begin{eqnarray}
    \psi(m\,\Delta t,n\,\Delta x) & \equiv & \psi_{m,n} \comma \\
    V(n\,\Delta x) & \equiv & V_n \comma \\
    u(n\,\Delta x) & \equiv & u_n  \comma
  \end{eqnarray}
  where the mesh size has to satisfy the condition
  $\Delta x > \Delta t$ so that the numerical rate of propagation
  of data is greater than its analytical counterpart. The
  discretization of the Cauchy problem above then implies
  \begin{eqnarray}
    \psi_{\minus 1,n} & = & \minus \Delta t\,u_n \comma \\
    \psi_{0,n} & = & 0 \comma \\
    \psi_{m+1,n} & = &
    \SqBk{2-2\,\frac{{\Delta t}^2}{{\Delta x}^2}-{\Delta t}^2\,V_n}\,\psi_{m,n}
    - \psi_{m-1,n}
    + \frac{{\Delta t}^2}{{\Delta x}^2}\,\SqBk{\psi_{m,n-1}+\psi_{m,n+1}}
    \period
  \end{eqnarray}
  As a result, we can follow the evolution of the field $\psi$
  starting from the initial data given at time $t = 0$.

  In the case where the black hole geometry is asymptotically
  flat, the tortoise coordinate $x$ goes from negative infinity to
  positive infinity. Therefore our Cauchy problem is similar to
  the infinite string problem in which the initial data propagates
  towards left and right indefinitely. The initial data no longer
  enjoys this privilege when the background is asymptotically
  anti-de Sitter because the tortoise coordinate goes from minus
  infinity to zero only. In other words, the right-propagating
  data cannot travel in this direction forever. Analogous to the
  semi-infinite vibrating string problem, boundary conditions at
  spatial infinity (i.e. $x = 0$) are needed in the asymptotically
  anti-de Sitter background in order to formulate the problem
  appropriately. Two types of boundary conditions that are widely
  used in anti-de Sitter backgrounds are the Dirichlet and Neumann
  conditions \cite{Lifschytz}. In our case, the former reads
  \begin{eqnarray}
    \psi(t,x=0) \; = \; 0
    \hspace{1cm} & {\rm and} & \hspace{1cm}
    \di_x \psi(t,x=0) \; = \; 1 \label{Dirichlet}
  \end{eqnarray}
  whilst the latter is simply
  \begin{eqnarray}
    \psi(t,x=0) \; = \; 1
    \hspace{1cm} & {\rm and} & \hspace{1cm}
    \di_x \psi(t,x=0) \; = \; 0 \period \label{Neumann}
  \end{eqnarray}
  We shall consider employing both of these boundary conditions 
  at spatial infinity for our
  numerical computations whenever the background geometry is
  asymptotically anti-de Sitter.

\section{Asymptotically Flat Backgrounds}
  In this section we will review the behaviour of radiative
  falloff in asymptotically flat background spacetimes
  \cite{Price}. We will present the results of the numerical
  calculation first.

\begin{figure}[htbp]
    \hfill \psfig{file=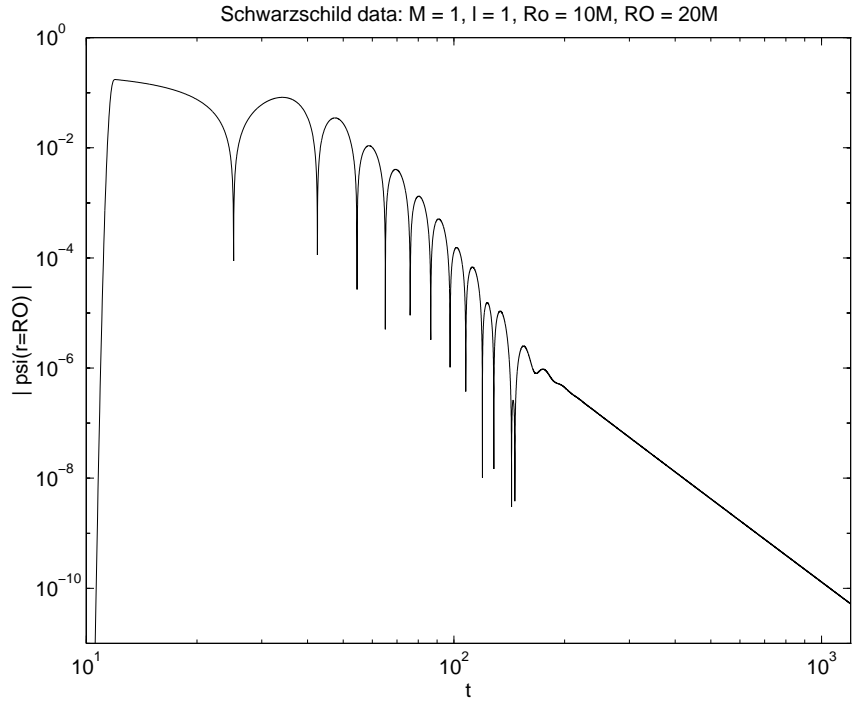,height=10cm} \hfill \mbox{}
    \caption{The decay of a scalar wave in Schwarzschild background.
	     Prior to $t=200$ the decay is accompanied by `ringing' of
	     the quasi-normal modes, after which the falloff rate is
	     that of an inverse power-law.}
    \label{Schwz1}
  \end{figure}

  Figure \ref{Schwz1} shows a sample inverse power-decay of a
  scalar wave in the Schwarzschild background spacetime of mass
  $M$. We solve the wave equation (\ref{WE5}) numerically using
  the scheme discussed in the previous section. The compact
  initial Gaussian impulse is centered at a distance $r = 10\,M$
  (or $x = 12.76\, M$) and for simplicity we choose the $l=1$
  spherical harmonic. Figure \ref{Schwz1} shows how the magnitude
  of the scalar field $\psi$ at a distance $r = 20\,M$ (i.e.
  $x = 24.40\, M$) evolves. Using linear regression, we find that
  the slope of the straight line on the graph is $\minus 5.026$ in
  agreement with the analytic prediction of an inverse power-law
  falloff with exponent $2\,l+3$ \cite{Price}. 

\begin{figure}[htbp]
    \hfill \psfig{file=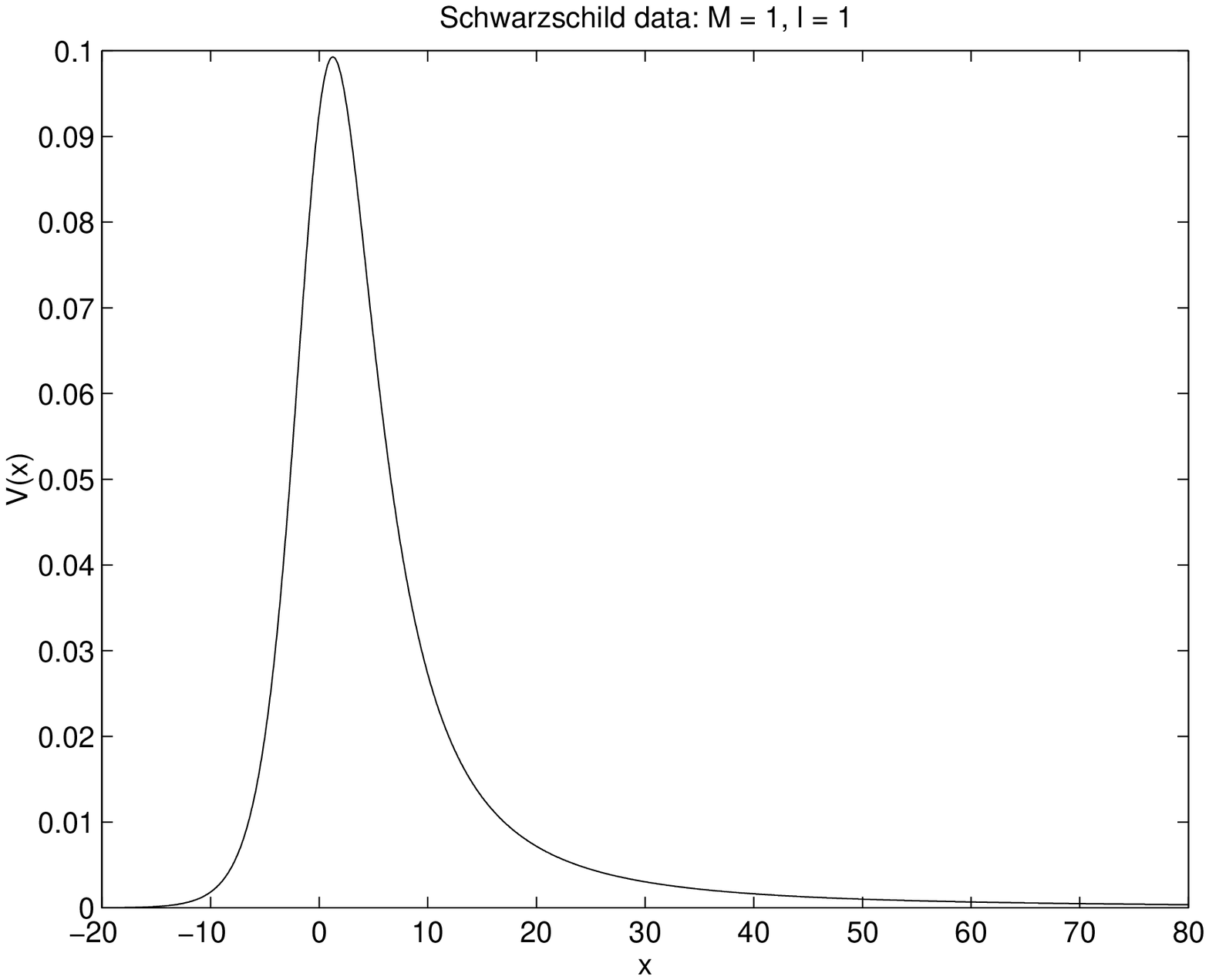,height=10cm} \hfill \mbox{}
    \caption{Potential barrier for the Schwarzschild background}
    \label{Schwz2}
  \end{figure}

\begin{figure}[htbp]
    \hfill \psfig{file=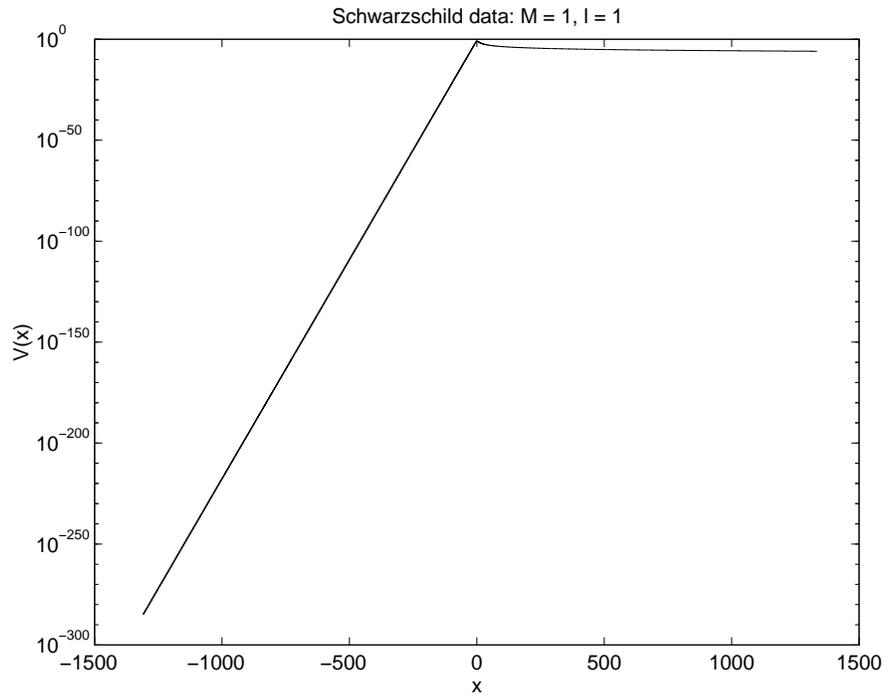,height=10cm} \hfill \mbox{}
    \caption{Exponential decrease on the left side of the barrier in $V(x)$}
    \label{Schwz3}
  \end{figure}

\begin{figure}[htbp]
    \hfill \psfig{file=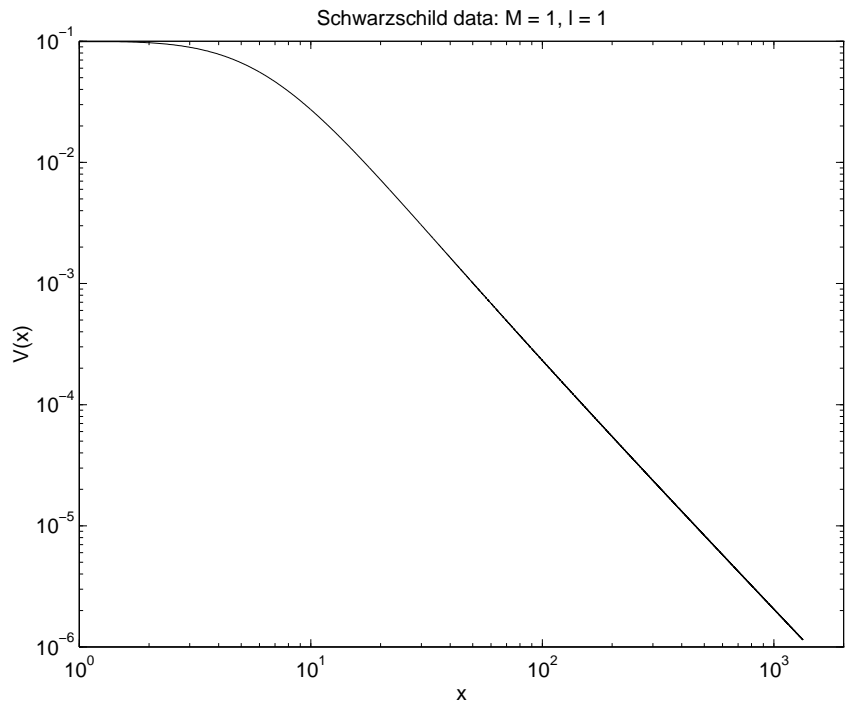,height=10cm} \hfill \mbox{}
    \caption{Inverse power decrease on the right side of the barrier of $V(x)$}
    \label{Schwz4}
  \end{figure}

\begin{figure}[htbp]
    \hfill \psfig{file=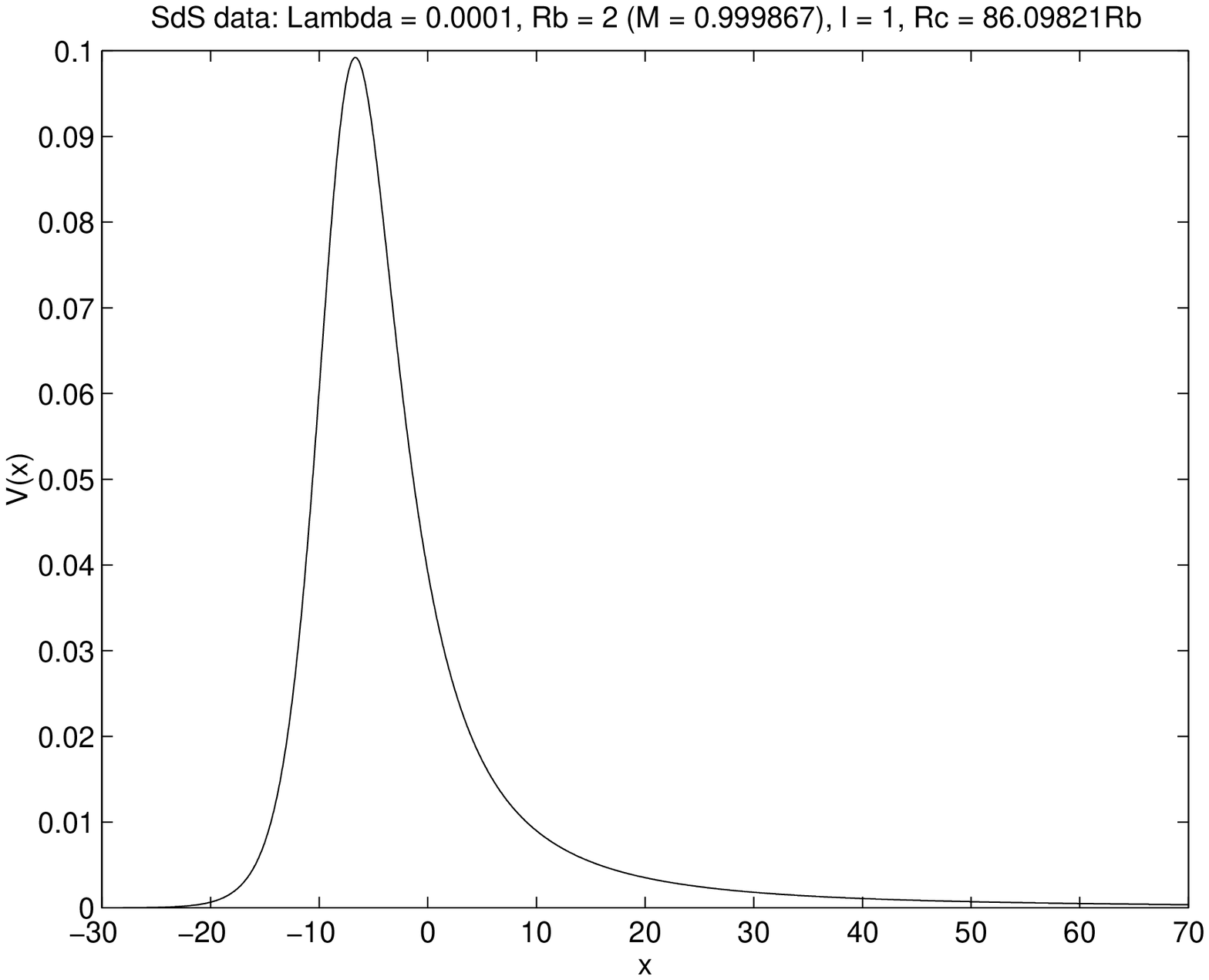,height=10cm} \hfill \mbox{}
    \caption{Potential barrier for the Schwarzschild-de Sitter background}
    \label{SdS1}
  \end{figure}

\begin{figure}[htbp]
    \hfill \psfig{file=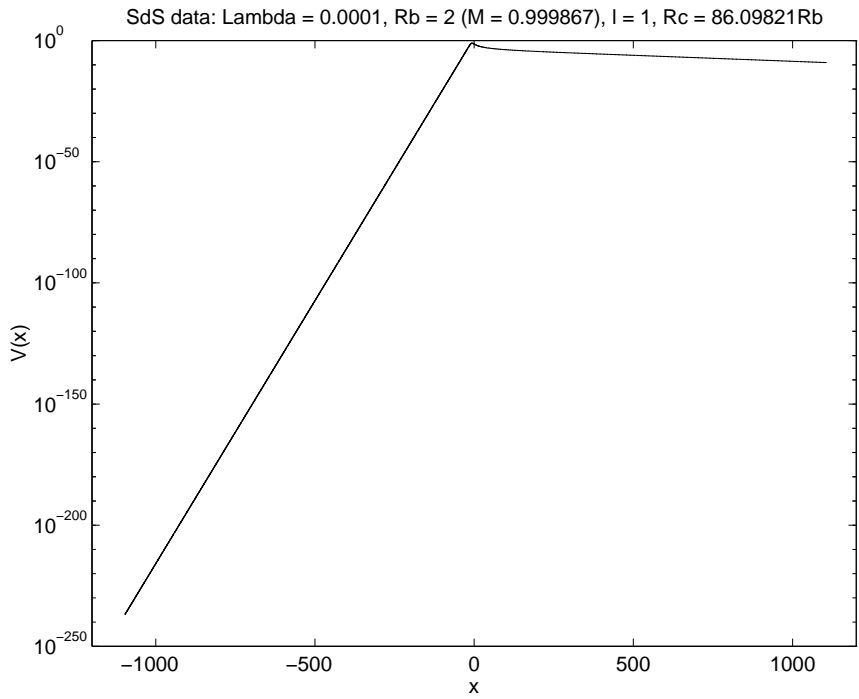,height=10cm} \hfill \mbox{}
    \caption{Exponential decreasing nature on the left side of
	     $V(x)$ in SdS background}
    \label{SdS2}
  \end{figure}

\begin{figure}[htbp]
    \hfill \psfig{file=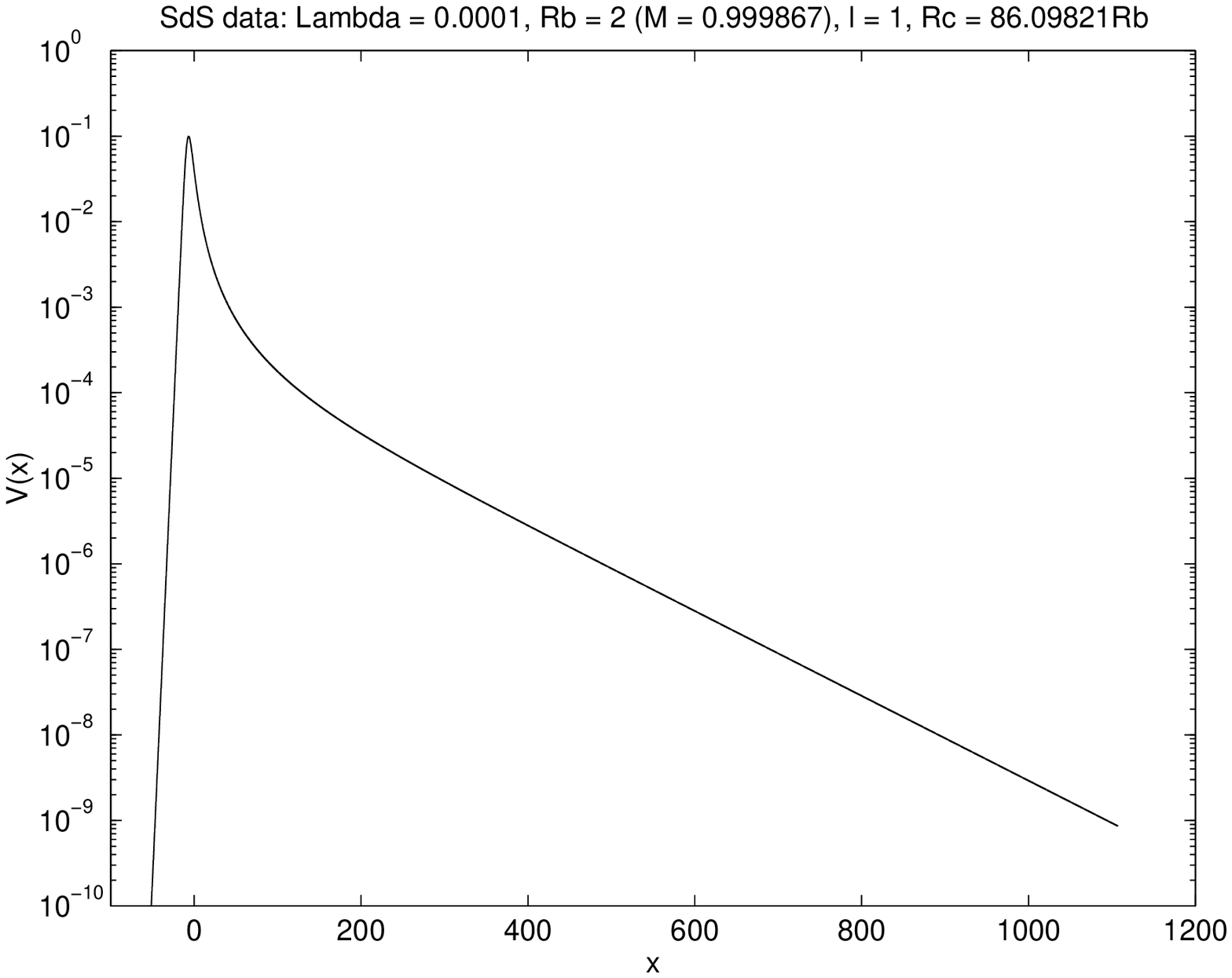,height=10cm} \hfill \mbox{}
    \caption{Exponential decrease of the right side of $V(x)$ in SdS background}
    \label{SdS3}
  \end{figure}

  Figures \ref{Schwz2} to \ref{Schwz4} show the same potential
  barrier $V(x)$ that is responsible for this falloff behaviour.
  Figure \ref{Schwz3} shows an exponential decrease of the left
  side of the potential function; this is a result of the fact
  that the event horizon is located at $x = \minus \infty$. On the
  other hand, the power-law decrease on the right side of $V(x)$
  shown in figure \ref{Schwz4} is a direct consequence of the use
  of an asymptotically flat background. If we compare figures
  \ref{Schwz2} to \ref{Schwz4} with figures \ref{SdS1} to
  \ref{SdS3} which represent the potential function in
  Schwarzschild-de Sitter (SdS) spacetime, we find that the right
  side of $V(x)$ has different decaying behaviour even the overall
  appearance of $V(x)$ on the linear graph is very similar. It is
  this difference that distinguishes the falloff behaviour in the
  two backgrounds

  Now let us consider a scalar wave in an asymptotically flat
  $(D+1)$-dimensional background. For the remainder of this
  section we will restrict our attention to the case where the
  number of spatial dimensions $D$ is odd. The motivation for this
  may be traced back to Huygen's Principle, which implies that in
  even spatial dimensions the scalar wave obeying the equation
  $\Del^2 \Psi = 0$ always develops a tail, regardless of whether
  or not the asymptotically flat background is sourceless.
  Consequently identification of the tail part of the wave that is
  due to solely to backscattering becomes quite problematic when
  $D$ is even.

  Inspired by the work of Ching et. al. \cite{Ching1,Ching2}, we
  consider a spherically symmetric static metric of the form
  (\ref{MetricForm}) with lapse function 
  \begin{eqnarray}
    N(r) & = & 1 - m\,\frac{\RdBk{\ln\!|r|}^\beta}{r^\alpha}
    \period \label{DLapse}
  \end{eqnarray}
  The constants $\alpha$ and $\beta$ are integers, where
  $\beta \ge 0$ but $\alpha > 0$. The other constant $m$ is a real
  number. When we have $\alpha = 1$, $\beta = 0$ and $D = 3$, this
  becomes a Schwarzschild background of mass $m/2$. Once we have
  the lapse function, we can compute the potential function 
  \begin{eqnarray}
    V_e(r) & = &
    \frac{(2\,l+D-3)\,(2\,l+D-1)}{4\,r^2}
    + m\,\frac{4\,\xi\,(D-1-\alpha)\,(D-2-\alpha)-(D-1)\,(D-3-2\,\alpha)}{4}\,\frac{\RdBk{\ln\!|r|}^\beta}{r^{\alpha+2}} \nonumber \\&&
    + \; m\,\beta\,\frac{2\,\xi\,(2\,D-3-2\,\alpha)-D+1}{2}\,\frac{\RdBk{\ln\!|r|}^{\beta-1}}{r^{\alpha+2}}
    + m\,\beta\,(\beta-1)\,\xi\,\frac{\RdBk{\ln\!|r|}^{\beta-2}}{r^{\alpha+2}}
  \end{eqnarray}
  from (\ref{Ve}). 

  We first find the static solution $\psi_S(r)$ of wave equation
  (\ref{WE2}), by obtaining the solution of the equation
  $\SOL{\psi_S(r)}=0$. It is straightforward to show that
  $\psi_S(r)$ has the following form:
  \begin{eqnarray}
    \psi_S(r) & = &
    r^{\minus \gamma}\,\sum_{j=0}^\infty \frac{a_j(r)}{r^{j\,\alpha}}
    + r^{\gamma+1}\,\sum_{j=0}^\infty \frac{c_j(r)}{r^{j\,\alpha}}
    \comma \label{DStatic}
  \end{eqnarray}
  where $\gamma \equiv l+(D-3)/2$. Notice that $\gamma$ is an
  integer when the spatial dimension $D$ is odd. Except when $D=1$
  this integer is always positive. Hence it is the first sum of
  the solution that is physically relevant since it vanishes for
  large $r$, and we choose this as the static solution. For the
  remainder of this section, we assume that $D\geq 3$ so that this
  choice of $\psi_S(r)$ is valid. The coefficients $a_0(r)$ and
  $c_0(r)$ are arbitrary constants but the other coefficients are
  all polynomial in $\ln\!|r|$. The generating equations for
  $a_j(r)$ and $c_j(r)$ are given in Appendix \ref{D+1apdx}.

  We follow the approach in refs. \cite{Price,Gundlach} and let
  \begin{eqnarray}
    \psi_I \; = \;
    \sum_{i=0}^\infty B_i(r)
    \SqBk{g^{(\minus i)}(u)+(\minus 1)^i\,f^{(\minus i)}(v)} \label{InitialForm}
  \end{eqnarray}
  be the form of the initial wave, i.e. the wave emitted by a star
  at the onset of gravitational collapse. In other words, this is
  the time when $t \ll r$. The functions $g(u)$ and $f(v)$ are as
  yet unknown. The term $g^{(\minus i)}(u)$ represents $i$
  integrations of the function $g(u)$ with respect to $u$;
  similarly for $f^{(\minus i)}(v)$. Using (\ref{InitialForm}),
  equation (\ref{WE4}) becomes
  \begin{eqnarray}
    0 & = &
    \half\,N(r)\,\frac{d}{d\,r} B_0(r)\,\SqBk{g^{(1)}(u)-f^{(1)}(v)}
    \nonumber \\ &&
    - \, \quarter\,N(r)\,\sum_{i=0}^\infty
    \BrBk{\SOL{B_i(r)}-2\,\frac{d}{d\,r} B_{i+1}(r)}\,\SqBk{g^{(\minus i)}(u)+(\minus 1)^i\,f^{(\minus i)}(v)} \period
  \end{eqnarray}
  This equation has a set of solutions
  \begin{eqnarray}
    B_0(r) & = & 1 \comma \label{B0} \\
    B_{i+1}(r) & = &
    \half\,N(r)\,\frac{d}{d\,r} B_i(r) - \half\,\int V_e(r)\,B_i(r)\,dr \comma
    \hspace{1cm} i = 0,1,2,\cdots \label{Bi}
  \end{eqnarray}
  where we have set $B_0(r)=$ constant $=1$ without loss of
  generality. The pair of equations above allow us to generate
  $B_i(r)$ hierarchically in a straightforward manner.

  We can split each $B_i(r)$ into two parts, denoted by $B_i^P(r)$
  and $B_i^T(r)$. $B_i^P(r)$ is defined as the $m$-independent
  portion of $B_i(r)$, while $B_i^T(r)$ is the rest which is
  $m$-dependent. Physically, the part $B_i^P(r)$ represents the
  wave on the lightcone because it is the part that would be
  generated if the background were flat ($m = 0$). Price referred
  to this part as the primary wave which depends only on the mode
  of the spherical harmonics. The other part $B_i^T(r)$ is called
  the tail of the wave because it is created by the presence of
  the spacetime curvature and is off the lightcone due to
  scattering. Given equations (\ref{B0}) and (\ref{Bi}), one can
  show that the primary part of $B_i(r)$ is simply
  \begin{eqnarray}
    B_i^P(r) & = &
    \frac{\Gamma(\gamma+1+i)}{2^i\,i!\,\Gamma(\gamma+1-i)\,r^i} \period
  \end{eqnarray}
  When $D$ is odd, $\gamma$ is an integer and the sequence
  $\{\,B_i^P(r)\,\}_{i=0}$ truncates. However if $D$ is even, the
  sequence does not terminate and the primary part of the initial
  wave $\psi_I$ has infinitely many terms. Indeed when this is the
  case, it is inappropriate to call $B_i^P(r)$ the primary part
  because there are tails present. The approach breaks down
  because the primary and tail parts of the wave become
  indistinguishable. This is the reason we restrict ourselves to
  odd $D$ in this section as mentioned earlier.

  Let us now consider the tail part of $B_i(r)$. Unlike the
  primary part, $B_i^T(r)$ has no simple solution. Fortunately one
  can always generate $B_i(r)$ recursively. Note that $B_i^T(r)$
  is of order $O(\,\RdBk{\ln\!|r|}^\beta/r^{i+\alpha}\,)$.

  Since $\gamma$ is an integer in odd spatial dimensions, it is
  convenient to define the functions $G(u)$ and $F(v)$ as
  \begin{eqnarray}
    G(u) \; = \; g^{(\minus \gamma)}(u) \hspace{1cm} {\rm and} \hspace{1cm}
    F(v) \; = \; f^{(\minus \gamma)}(v)
  \end{eqnarray}
  such that the initial wave can be expressed as
  $\psi_I = \sum B_i(r)\,{T_I}^{\gamma-i}(u,v)$ where
  \begin{eqnarray}
    {T_I}^{\gamma-i}(u,v) & \equiv &
    G^{(\gamma-i)}(u) + (\minus 1)^i\,F^{(\gamma-i)}(v) \period
  \end{eqnarray}
  When the background is flat ($m = 0$), only the first
  $\gamma + 1$ terms of $B_i^P(r)$ survive. We expect that any
  outgoing radiation will propagate to spatial infinity without
  any scattering in a flat background. Therefore we must have
  \begin{eqnarray}
    F^{(\gamma)}(v) \; = \; F^{(\gamma-1)}(v) \; = \; \cdots \; = \;
    F^{(1)}(v) \; = \; F^{(0)}(v) \; = \; 0 \label{InitF}
  \end{eqnarray}
  because $F(v)$ represents the scattered infalling radiation and
  the $B_i^P(r)$ are non-zero only for $0 \le i \le \gamma$. In
  other words, we insist that $F(v) = 0$.

  For concreteness, we suppose the scalar wave starts leaving the
  star at retarded time $u=U_0$; that is to say the scalar field
  is zero before $u=U_0$. Continuity then implies
  \begin{eqnarray}
    G^{(\gamma)}(U_0) \; = \; G^{(\gamma-1)}(U_0) \; = \; \cdots \; = \;
    G^{(1)}(U_0) \; = \; G^{(0)}(U_0) \; = \; 0 \label{InitG1} \\
    G^{(\minus 1)}(u) \; = \; \int^u_{U_0} G^{(0)}(\zeta)\,d\zeta \comma
    \hspace{1cm}
    G^{(\minus 2)}(u) \; = \; \int^u_{U_0} G^{(\minus 1)}(\zeta)\,d\zeta
    \comma \hspace{1cm} \cdots \label{InitG2}
  \end{eqnarray}
  We also assume that there is essentially no emission from the
  star after $u = U_1 > U_0$ (due to gravitational red shift).
  When the star possesses no initial static moment at the onset of
  collapse, this cutoff condition is translated into
  \begin{eqnarray}
    G^{(\gamma)}(u) \; = \; G^{(\gamma-1)}(u) \; = \; \cdots \; = \;
    G^{(1)}(u) \; = \; G^{(0)}(u) \; = \; 0 \comma
    \hspace{1cm} \forall \; u \ge U_1 > U_0 \label{CutoffNSM}
  \end{eqnarray}
  because only the terms $G(u)$ to $G^{(\gamma)}(u)$ which
  represent the primary wave correspond to the emission from the
  star. This condition implies
  \begin{eqnarray}
    G^{(\minus 1)}(u) \; = \;
    \int^u_{U_0} G^{(0)}(\zeta)\,d\zeta \; = \;
    \int^{U_1}_{U_0} G^{(0)}(\zeta)\,d\zeta \; = \;
    {\rm constant} \comma \hspace{1cm} \forall \; u \ge U_1 \period
  \end{eqnarray}
  Therefore if there is no static moment before $u = U_0$, the
  scalar wave after time $u = U_1$ is
  \begin{eqnarray}
    \psi_{u \ge U_1} & = &
    \sum_{i=\gamma+1}^\infty B_i(r)\,{T_I}^{\gamma-i}(u,v) \comma
  \end{eqnarray}
  where ${T_I}^{\minus 1}(u,v)$ is only a constant and
  $B_i(r) = B_i^T(r)$ is of order $O(\,\RdBk{\ln\!|r|}^\beta/r^{i+\alpha}\,)$.
  So the first term of this solution which is time-independent is
  of order $O(\,\RdBk{\ln\!|r|}^\beta/r^{\gamma+\alpha+1}\,)$.
  Physically, this says the tail of the perturbation persists
  after the time $u = U_1$. The primary part of the wave has been
  ``washed out'' by the gravitational red shift before $u = U_1$.

  The situation is more complicated if the star carries a static
  moment at the onset of the collapse because the static wave
  $\psi_S(r)$ can also be divided into primary and tail parts. In
  this case, we superimpose $\psi_S(r)$ of (\ref{DStatic}) and
  $\psi_I(r)$ of (\ref{InitialForm}) together to form a new
  initial scalar wave. We also suppose that only the tail of the
  perturbation persists after the retarded time $u = U_1$
  \cite{Price}. For the superimposed initial wave   
  this cutoff condition   requires
  \begin{eqnarray}
    G^{(0)}(u) & = & \minus \frac{2^\gamma\,\gamma!}{(2\,\gamma)!} \comma
    \hspace{1cm} \forall \; u \ge U_1 \ge U_0 \label{CutoffWSM}
  \end{eqnarray}
  which yields
  \begin{eqnarray}
    G^{(\gamma)}(u) \; = \; G^{(\gamma-1)}(u) \; = \; \cdots \; = \;
    G^{(1)}(u) \; = \; 0 \period
  \end{eqnarray}
  The restriction on $G(u)$ above ensures that the primary part of
  (\ref{InitialForm}) can be cancelled by the $m$-independent part
  of (\ref{DStatic}) properly after the time $U_1$, leaving the
  combined $\psi_I$ to be $m$-dependent only after $U_1$. As a
  result, the wave with initial static moment will become
  \begin{eqnarray}
    \psi_{u \ge U_1} & = &
    \frac{1}{r^\gamma}\,\sum_{j=1}^\infty \frac{a_j(r)}{r^{j\,\alpha}}
    + \sum_{i=\gamma}^\infty B_i^T(r)\,{T_I}^{\gamma-i}(u,v) \period
  \end{eqnarray}
  One can show that the first sum in the equation above is of
  order $O(\,\RdBk{\ln\!|r|}^\beta/r^{\gamma+\alpha}\,)$ because
  the coefficient function $a_1(r)$ is a polynomial in $\ln\!|r|$
  with degree $\beta$. In the second sum, the term
  ${T_I}^{0}(u,v)$ is a constant which equals to $G(u)$, and
  $B_\gamma^T(r)$ is of order $O(\,\RdBk{\ln\!|r|}^\beta/r^{\gamma+\alpha}\,)$.
  Therefore instead of having a time independent part of order
  $O(\,\RdBk{\ln\!|r|}^\beta/r^{\gamma+\alpha+1}\,)$, the wave
  with initial static moment has an order
  $O(\,\RdBk{\ln\!|r|}^\beta/r^{\gamma+\alpha}\,)$ only.

  Now we turn our attention to the scalar wave at late time. By
  late time, we mean $t \gg r$. For the late time wave $\psi_L$,
  we introduce another ansatz
  \begin{eqnarray}
    \psi_L \; = \;
    \sum_{i=0}^\infty C_i(r)\,T^i_L(u,v) \comma
    & \hspace{2cm} &
    T^i_L(u,v) \; \equiv \;
    I^{(i)}(u) + (\minus 1)^i\,H^{(i)}(v)
    \period
  \end{eqnarray}
  By substituting $\psi = \psi_L$, equation (\ref{WE4}) becomes
  \begin{eqnarray}
    \quarter\,N(r)\,\SOL{C_0(r)}\,T^0_L(u,v)
    + \quarter\,N(r)\,\sum_{i=0}^\infty \BrBk{\SOL{C_{i+1}(r)}-2\,\frac{d}{d\,r}
 C_i(r)}\,T^{i+1}_L(u,v)
    & = & 0 \period
  \end{eqnarray}
  This equation has another set of solutions
  \begin{eqnarray}
    \SOL{C_0(r)} & = & 0 \comma \label{diffC0} \\
    \SOL{C_{i+1}(r)} & = & 2\,\frac{d}{d\,r} C_i(r)
    \comma \hspace{1cm} i = 0,1,2,\cdots \label{diffCi}
  \end{eqnarray}
  Unlike the case for the initial wave $\psi_I$, (for which the
  functions $B_i(r)$ can be calculated recursively using
  straightforward differentiation and integration), recursive
  generation of the functions $C_i(r)$ involve inverting the
  differential operator ${\cal L}$. The zero order equation has a
  solution
  \begin{eqnarray}
    C_0(r) & = &
    r^{\gamma+1}\,\sum_{j=0}^\infty \frac{c_j(r)}{r^{j\,\alpha}} \period
  \end{eqnarray}
  The coefficient functions $c_j(r)$ are those in equations
  (\ref{DStatic}). They can be calculated by using the generating
  equations in Appendix \ref{D+1apdx}. The other inhomogeneous
  differential equation (\ref{diffCi}) has a solution of the form
  \begin{eqnarray}
    C_i(r) & = &
    r^{\gamma+1+i}\,\sum_{j=0}^\infty \frac{c^i_j(r)}{r^{j\,\alpha}} \comma
  \end{eqnarray}
  where $i = 0, 1, 2, \cdots$ When $i = 0$, it is understood that
  $c^0_j(r) = c_j(r)$. The coefficients $c^i_j(r)$ are also given
  in Appendix \ref{D+1apdx}. Since each coefficient $c^i_0(r)$ is
  a constant instead of a polynomial in $\ln\!|r|$, we can estimate
  the order of the late time wave $\psi_L$ as
  \begin{eqnarray}
    \psi_L & = &
    \sum_{i=0}^\infty O\!\RdBk{r^{\gamma+1+i}}\,{T_L}^i(u,v) \period
  \end{eqnarray}

  Finally we match the late time solution to the initial solution
  at some transient period where $u$, $v$ and $r$ are of the same
  order. As the background is asymptotically flat, the tortoise
  coordinate $x$ must have an order similar to that of $r$ in this
  transient region. That is to say the orders of $r$ and $t$ are
  the same. The initial wave $\psi_{u \ge U_1}$ in this period
  becomes
  \begin{eqnarray*}
    \psi_{u \ge U_1} \; \sim \;
    O\!\RdBk{\frac{\RdBk{\ln\!|r|}^\beta}{r^{\gamma+\alpha+1}}} \comma
    & \hspace{1cm} &
    {\rm without \;\; initial \;\; static \;\; moment} \comma \\
    \psi_{u \ge U_1} \; \sim \;
    O\!\RdBk{\frac{\RdBk{\ln\!|r|}^\beta}{r^{\gamma+\alpha}}} \comma
    & \hspace{1cm} &
    {\rm with \;\; initial \;\; static \;\; moment} \period
  \end{eqnarray*}
  In order to have consistent orders in the transient period, we
  must have
  \begin{eqnarray}
    {T_L}^0(t,t) \; \sim \;
    O\!\RdBk{\frac{\RdBk{\ln\!|t|}^\beta}{t^{2\,\gamma+2+\alpha}}}
    \label{rate1}
  \end{eqnarray}
  if the star has no initial static moment; whilst the order of
  ${T_L}^0$ must be
  \begin{eqnarray}
    {T_L}^0(t,t) \; \sim \;
    O\!\RdBk{\frac{\RdBk{\ln\!|t|}^\beta}{t^{2\,\gamma+1+\alpha}}}
    \label{rate2}
  \end{eqnarray}
  if the star has a static moment at the onset of collapse. In
  other words equations (\ref{rate1}) and (\ref{rate2}) give the
  falloff behaviour of the wave at late time. The inverse power
  falloff behaviour modified by a logarithmic term was first noted
  by Ching {\sl et al.} \cite{Ching2}. For a Schwarzschild
  background, we set $D = 3$, $\alpha = 1$ and $\beta = 0$ and
  have $\gamma = l$, yielding the familiar power-law decay rate
  \cite{Price,Gundlach}.

\section{The 3D Black Hole Background}
  The $(2+1)$-dimensional black hole spacetime obtained by Banados
  Teitelboim and Zanelli \cite{BTZ} is a spacetime which satisfies
  the vacuum Einstein equations with a negative cosmological
  constant ($\Lambda < 0$). Its metric is
  \begin{eqnarray}
    ds^2 & = &
    \minus N(r)\,dt^2 + \frac{dr^2}{N(r)}
    + r^2\,\RdBk{\minus \frac{J}{2\,r^2}\,dt + d\phi}^2 \comma \\
    N(r) & = & |\Lambda|\,r^2 - M + \frac{J^2}{4\,r^2}
  \end{eqnarray}
  and is actually anti-de Sitter spacetime with identifications.
  The constant $M > 0$ is the quasi-local mass of the black hole
  and $J$ is the angular momentum of the hole \cite{BCM}.
  For our purposes we shall regard the above metric as a spacetime
  which is asymptotically anti-de Sitter.

  When $J = 0$, the wave equation (\ref{WE1}) gives (\ref{WE5})
  with
  \begin{eqnarray}
    V_e(r) & = &
    \frac{3\,|\Lambda|}{4}\,(1-8\,\xi) + \frac{M+4\,l^2}{4\,r^2}
    \period \label{BTZVe}
  \end{eqnarray}
  In this case, the tortoise coordinate $x$ is given by
  \begin{eqnarray}
    x & \equiv &
    \int \frac{dr}{N(r)} \; = \;
    \frac{1}{2\,\sqrt{|\Lambda|\,M}}\,\ln\!\left|\,\frac{\sqrt{|\Lambda|}\,r-\sqrt{M}}{\sqrt{|\Lambda|}\,r+\sqrt{M}}\,\right| \period
  \end{eqnarray}
  As $r$ goes from $\sqrt{M/|\Lambda|}$ to infinity, the tortoise
  coordinate has a range $\RdBk{\minus \infty\,,\,0}$. Thus we can
  write $r$ in terms of $x$ as
  \begin{eqnarray}
    r & = &
    \sqrt{\frac{M}{|\Lambda|}}\,\frac{1+\exp(2\,\sqrt{|\Lambda|\,M}\,x)}{1-\exp(2\,\sqrt{|\Lambda|\,M}\,x)} \period
  \end{eqnarray}
  For conformal wave in $2+1$ dimensions, the parameter $\xi$
  equals $1/8$ and the potential barrier $V$ becomes
  \begin{eqnarray}
    V(x) \; = \; V_0\,\frac{\exp(2\,\lambda\,x)}{\RdBk{1+\exp(2\,\lambda\,x)}^2}
    \; \approx \; V_0\,\exp(2\,\lambda\,x)
    \comma \label{BTZpotential} \label{Vapprox}
  \end{eqnarray}
  where, for convenience, we have defined
  \begin{eqnarray}
    V_0 \; \equiv \; |\Lambda|\,(M + 4\,l^2) \; > \; 0
    \hspace{1cm} {\rm and} \hspace{1cm}
    \lambda \; \equiv \; \sqrt{|\Lambda|\,M} \; > \; 0 \period
  \end{eqnarray}

  The procedure for finding a solution to the conformal
  scalar wave equation is as follows. 
  We first write down the representation
  of the solution $\psi$ in terms of Green's functions. The rest
  of the problem then reduces to that of looking for the correct
  Green's function. We will Fourier transform from the time domain
  to the frequency domain and obtain the Green's function in
  frequency space. Once we have the Green's function in frequency
  domain, an inverse Fourier transformation will yield the
  solution $\psi(t,x)$. The solution is exact for the approximate
  potential (\ref{Vapprox}). We will then use this exact solution
  and the Born approximation to compute the next correction of the
  solution for a more accurate description of the potential. A
  case when $J \ne 0$ will be studied as well.

  We first assume that there exists a Green's function
  \begin{eqnarray}
    G(x,\xi;t-\tau) & = & G(\xi,x;t-\tau)
  \end{eqnarray}
  which is zero when $t < \tau$. We define an operator $D$ as
  \begin{eqnarray}
    D & = & \di_{tt} - \di_{xx} + V(x) \comma
  \end{eqnarray}
  such that the Green's function with respect to this operator has
  the property
  \begin{eqnarray}
    D G(x,\xi;t-\tau) \; = \;
    \SqBk{\di_{tt} - \di_{xx} + V(x)}\,G(x,\xi;t-\tau) \; = \;
    \delta(t-\tau)\,\delta(x-\xi) \period \label{BTZDt}
  \end{eqnarray}
  The inner product between $D \psi(t,x)$ and $G(x,\xi;t-\tau)$
  gives
  \begin{eqnarray}
    \psi(t,x) & = &
    \int_{\minus \infty}^0 \SqBk{G(x,\xi;t)\,\di_t \psi(0,\xi)+\psi(0,\xi)\,\di_t G(x,\xi;t)}\,d\xi \nonumber\\&&
    + \; \int_0^\infty \SqBk{G(x,\xi;t-\tau)\,\di_{\xi} \psi(\tau,\xi)-\psi(\tau,\xi)\,\di_{\xi} G(x,\xi;t-\tau)}^{\xi=0}_{\xi=\minus \infty}\,d\tau
    \period \label{BTZpsi1}
  \end{eqnarray}
  As a result, we have changed the question from looking for
  $\psi(t,x)$ to searching for an appropriate Green's function.

  We now carry out a Fourier transformation and define
  \begin{eqnarray}
    \tilde{G}(x,\xi;\omega) & = &
    \int_{\minus \infty}^\infty G(x,\xi;t)\,\exp(i\,\omega\,t)\,dt \period
  \end{eqnarray}
  Therefore equation (\ref{BTZDt}) becomes
  \begin{eqnarray}
    \tilde{D} \tilde{G}(x,\xi;\omega) & = &
    \SqBk{\minus \omega^2 - \di_{xx} + V(x)}\,\tilde{G}(x,\xi;\omega) \; = \;
    \delta(x-\xi) \label{BTZDf}
  \end{eqnarray}
  if the Green's function $G(x,\xi;t)$ satisfies the conditions
  \begin{eqnarray}
    \lim_{t \rightarrow \infty} G(x,\xi;t) \; = \;
    \lim_{t \rightarrow \infty} \di_t G(x,\xi;t) \; = \; 0 \period
  \end{eqnarray}
  On physical grounds these assumptions are reasonable because any
  localized quantity is expected to be dispersed throughout the
  space by means of wave propagation. 
  Mathematically, these assumptions are consistent with the Fourier
  transformability of the function $G(x,\xi;t)$ which must be 
  absolutely integrable over $\Re$ in order for $\tilde{G}(x,\xi;\omega)$
  to be well-defined.   The Green's function in
  frequency space can be represented as
  \begin{eqnarray}
    \tilde{G}(x,\xi;\omega) & = &
    \left\{
    \begin{array}{cr}
      \frac{f(\xi;\omega)\,g(x;\omega)}{W(\omega;g,f)} & \hspace{2cm} {\rm if} \;\;\xi<x \comma
      \\\\
      \frac{f(x;\omega)\,g(\xi;\omega)}{W(\omega;g,f)} & \hspace{2cm} {\rm if} \;\;x<\xi \comma
    \end{array}
    \right.
  \end{eqnarray}
  where the function $W(\omega;g,f)$ is the Wronskian of two
  linearly independent functions $f(x;\omega)$ and $g(x;\omega)$,
  that is
  \begin{eqnarray}
    W(\omega;g,f) & = &
    g(x;\omega)\,\di_x f(x;\omega) - f(x;\omega)\,\di_x g(x;\omega) \period
  \end{eqnarray}
  The functions $f(x;\omega)$ and $g(x;\omega)$ are two independent
  solutions of the equations
  \begin{eqnarray}
    \tilde{D} f(x;\omega) \; = \; \tilde{D} g(x;\omega) \; = \; 0
  \end{eqnarray}
  so that $W(\omega;g,f)$ is independent on $x$.

  Let $\tilde{\psi}(x;\omega)$ be the Fourier transformation of
  the solution $\psi(t,x)$. We impose boundary conditions on the
  functions $f(x;\omega)$ and $g(x;\omega)$ so that
  \begin{eqnarray}
    f(x;\omega) \; \propto \; \tilde{\psi}(x;\omega)
    \hspace{1cm} & {\rm and} & \hspace{1cm}
    \di_x f(x;\omega) \; \propto \; \di_x \tilde{\psi}(x;\omega)
  \end{eqnarray}
  at the point $x \rightarrow \minus \infty$. At the other end
  ($x = 0$) we insist that
  \begin{eqnarray}
    g(x;\omega) \; \propto \; \tilde{\psi}(x;\omega)
    \hspace{1cm} & {\rm and} & \hspace{1cm}
    \di_x g(x;\omega) \; \propto \; \di_x \tilde{\psi}(x;\omega) \period
  \end{eqnarray}
  In equation (\ref{BTZpsi1}), after we have used the inverse
  Fourier transformation and interchanged the integrals with
  respect to $d\omega$ and $d\tau$, the representation of
  $\psi(t,x)$ simply becomes
  \begin{eqnarray}
    \psi(t,x) & = &
    \int_{\minus \infty}^0 \SqBk{G(x,\xi;t)\,\di_t \psi(0,\xi)+\psi(0,\xi)\,\di_t G(x,\xi;t)}\,d\xi
    \period \label{BTZpsi2}
  \end{eqnarray}

\subsection{Solution for an Approximate Potential}

  We will obtain the exact solution for the scalar wave $\psi$
  corresponding to the approximate potential barrier
  (\ref{Vapprox}) in a static 3D black hole background. 
  This entails finding the solution of the differential
  equation $\tilde{D} h(x;\omega) = 0$. This equation has two
  solutions \cite{Ching1}, namely
  \begin{eqnarray}
    h(x;\omega) & = & I_\nu\!\RdBk{z(x)}
    \hspace{1cm} {\rm or} \hspace{1cm} K_\nu\!\RdBk{z(x)} \comma \\
    z(x) & = & Z_0\,\exp(\lambda\,x) \; = \;
    \frac{\sqrt{V_0}}{\lambda}\,\exp(\lambda\,x) \comma
  \end{eqnarray}
  where $\nu \equiv \minus i\,\omega/\lambda$, and $I_\nu$ and
  $K_\nu$ are the modified Bessel functions of complex order $\nu$
  \cite{Abramowitz}. At the boundary $x \rightarrow \minus \infty$
  (the event horizon of the 3D black hole) we employ the condition
  \begin{eqnarray}
    f(x;\omega) & = & \exp(\minus i\,\omega\,x)
    \hspace{2cm} x \rightarrow \minus \infty
  \end{eqnarray}
  as usual \cite{Ching1}. Since this spacetime is asymptotically
  anti-de Sitter instead of flat, the boundary condition at
  spatial infinity is less trivial. Choosing Dirichlet conditions
  \cite{Lifschytz} 
  \begin{eqnarray}
    g(x=0;\omega) \; = \; 0 
    \hspace{1cm} & {\rm and} & \hspace{1cm}
    \di_x g(x;\omega) \vert_{x=0} \; = \; 1 \comma
  \end{eqnarray}
  the two functions $f(x;\omega)$ and $g(x;\omega)$ read
  \begin{eqnarray}
    f(x;\omega) & = &
    {\cal N}_f\,I_\nu\!\RdBk{z(x)}
    \comma \label{BTZf0} \\
    g(x;\omega) & = &
    {\cal N}_g\,\SqBk{I_\nu(Z_0)\,K_\nu\!\RdBk{z(x)} - K_\nu(Z_0)\,I_\nu\!\RdBk{z(x)}}
    \comma \label{BTZg0}
  \end{eqnarray}
  where the normalization coefficients ${\cal N}_f$ and
  ${\cal N}_g$ are given by
  \begin{eqnarray}
    {\cal N}_f \; = \;
    \RdBk{\frac{2}{Z_0}}^\nu\,\Gamma(1+\nu)
    \hspace{1cm} & {\rm and} & \hspace{1cm}
    {\cal N}_g \; = \; \minus \frac{1}{\lambda} \period
  \end{eqnarray}
  Now we put everything together and obtain
  \begin{eqnarray}
    \tilde{G}(x,\xi<x;\omega) & = &
    \frac{I_\nu\!\RdBk{z(\xi)}\,\SqBk{I_\nu(Z_0)\,K_\nu\!\RdBk{z(x)} - K_\nu(Z_0)\,I_\nu\!\RdBk{z(x)}}}{\lambda\,I_\nu(Z_0)}
    \period \label{BTZGreenf}
  \end{eqnarray}
  The remainder of the problem is to bring it from the frequency
  domain back to the time domain.

  The inverse Fourier transformation is given by the equation
  \begin{eqnarray}
    G(x,\xi;t) & = &
    \frac{1}{2\,\pi}\,\int^\infty_{\minus \infty} \tilde{G}(x,\xi;\omega)\,\exp(\minus i\,\omega\,t)\,d\omega \period
  \end{eqnarray}
  We evaluate this integral by analytically extending $\omega$ to
  complex values and using Cauchy's residue theorem. The contour
  of integration is chosen to be a large, closed semi-circle on
  the lower half $\omega$-plane, with center at $\omega = 0$. By
  Jordan's lemma, the contribution from the arc of the circle goes
  to zero as the radius of the arc tends to infinity. The Green's
  function then becomes
  \begin{eqnarray}
    G(x,\xi;t) & = &
    i\,\sum {\rm Res}\!\BrBk{\tilde{G}(x,\xi;\omega)\,\exp(\minus i\,\omega\,t)}
    \comma
  \end{eqnarray}
  where ${\rm Res}\!\BrBk{h(z)}$ denotes the residue of the
  function $h(z)$ at a pole. In our case, the poles of the Green's
  function $\tilde{G}(x,\xi;\omega)$ come from the term
  $I_\nu(Z_0)$ only. This term causes trouble because it has
  infinitely many zeros in the lower half $\omega$-plane. However
  $\nu \equiv \minus i\,\omega / \lambda = 0$ is not a root of
  $I_\nu(Z_0)$ for positive definite $Z_0$ and so $G(x,\xi;t)$
  decays to zero exponentially. The dominant exponential decaying
  rate is determined by the zero closest to the real axis on the
  lower half $\omega$-plane. In general this root has a
  non-vanishing real part which contributes to the oscillatory
  nature of the falloff behaviour. As a result, the exponentially
  decaying Green's function $G(x,\xi;t)$ implies that $\psi(t,x)$
  also decays in this manner.

  Had we chosen the Neumann condition instead of the Dirichlet
  condition for $g(x,\omega)$ \cite{Lifschytz}, that is
  \begin{eqnarray}
    g(x=0;\omega) \; = \; 1 \hspace{1cm} & {\rm and} & \hspace{1cm}
    \di_x g(x;\omega) \vert_{x=0} \; = \; 0 \comma
  \end{eqnarray}
  the function $g(x,\omega)$ would simply become
  \begin{eqnarray}
    g(x;\omega) & = &
    Z_0\,\SqBk{{I_\nu}'(Z_0)\,K_\nu\!\RdBk{z(x)} - {K_\nu}'(Z_0)\,I_\nu\!\RdBk{z(x)}} \period
  \end{eqnarray}
  The change in ${\cal N}_g$ does not affect the result in any way
  because it is always cancelled in $\tilde{G}$ when the Wronskian
  is taken into account. The change from $I_\nu(Z_0)$ to
  ${I_\nu}'(Z_0)$ causes the singular term in $\tilde{G}$ to
  become ${I_\nu}'(Z_0)$ which also has infinitely many roots on
  the lower half $\omega$-plane. Again, the origin is not a root.
  Therefore even if we employ the Neumann condition, the solution
  $\psi(t,x)$ also dies out to zero at an exponential rate.

\begin{figure}[htbp]
    \hfill \psfig{file=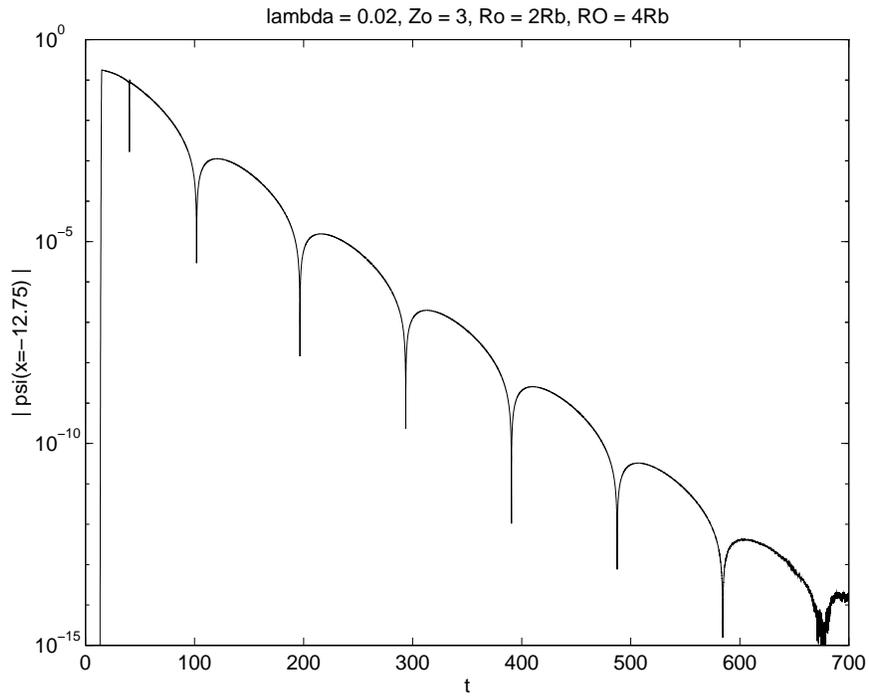,height=10cm} \hfill \mbox{}
    \caption{Exponential falloff for a conformal scalar wave using
	     the approximate potential (\ref{Vapprox}) and Dirichlet
	     boundary condition}
    \label{p_BTZ1}
  \end{figure}

\begin{figure}[htbp]
    \hfill \psfig{file=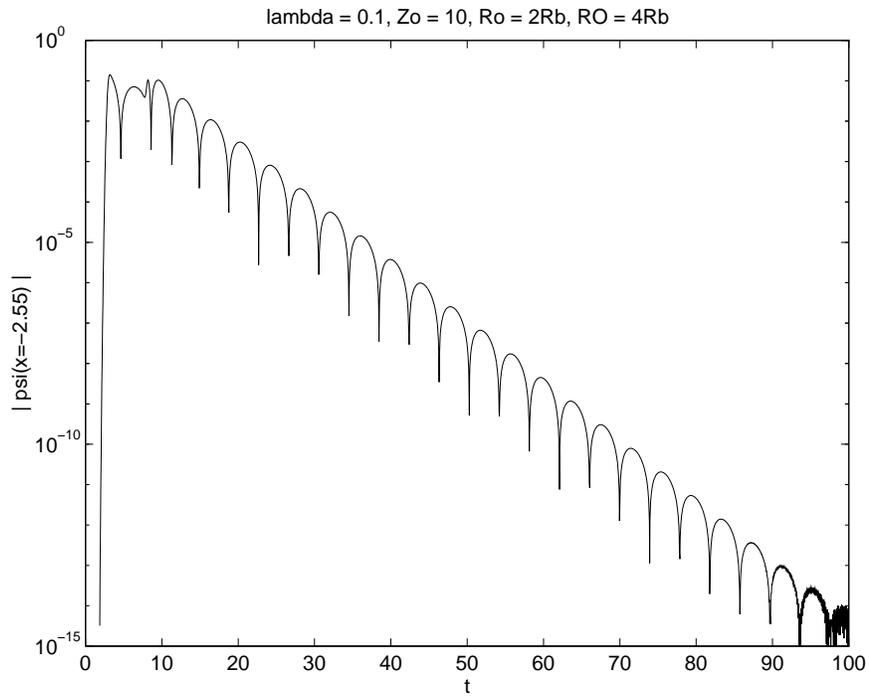,height=10cm} \hfill \mbox{}
    \caption{Another semilog graph as in figure \ref{p_BTZ1}
	     but with different parameters}
    \label{p_BTZ2}
  \end{figure}

  The next few graphs are the numerical results using the
  exponential potential. We used the same initial condition as in
  the numerical computation for the Schwarzschild case. Since the
  background is asymptotically anti-de Sitter, we integrate
  equation (\ref{WE5}) numerically with the Dirichlet boundary
  condition at $r = \infty$ (i.e. $x = 0$.) Figures \ref{p_BTZ1}
  and \ref{p_BTZ2} illustrate the falloff behaviour of the wave
  using different exponential potential functions. In both graphs,
  the initial Gaussian impulse is located at twice the black hole
  radius ($2\,R_b$) and the observation is made at $4\,R_b$. The
  straight line asymptote of the ringing behaviour on the semilog
  graph corresponds to exponential falloff, numerically confirming
  that the wave exponentially decays in this asymptotically
  non-flat background.

  A direct check of our analytic prediction against the numerical
  computation is problematic because no analytic procedure exists
  for computing the (complex) values of $\nu$ for which
  $I_\nu(Z_0)=0$ for fixed $Z_0$. Hence finding the root of
  $I_{\minus i\,\omega / \lambda}(Z_0)$ which is closest to the
  real axis on the lower half $\omega$-plane must be done
  numerically, by checking whether the root measured from the
  numerically generated graph (e.g. Figure \ref{p_BTZ1}) satisfies
  the equation $I_{\minus i\,\omega / \lambda}(Z_0) = 0$ for the
  relevant $Z_0$ (e.g. $Z_0=3$ for Figure \ref{p_BTZ1}). We have
  checked that the numerically determined roots satisfy this
  criterion for various values of $Z_0$ for all computations
  carried out in the $(2+1)$-dimensional case.


\begin{figure}[htbp]
    \hfill \psfig{file=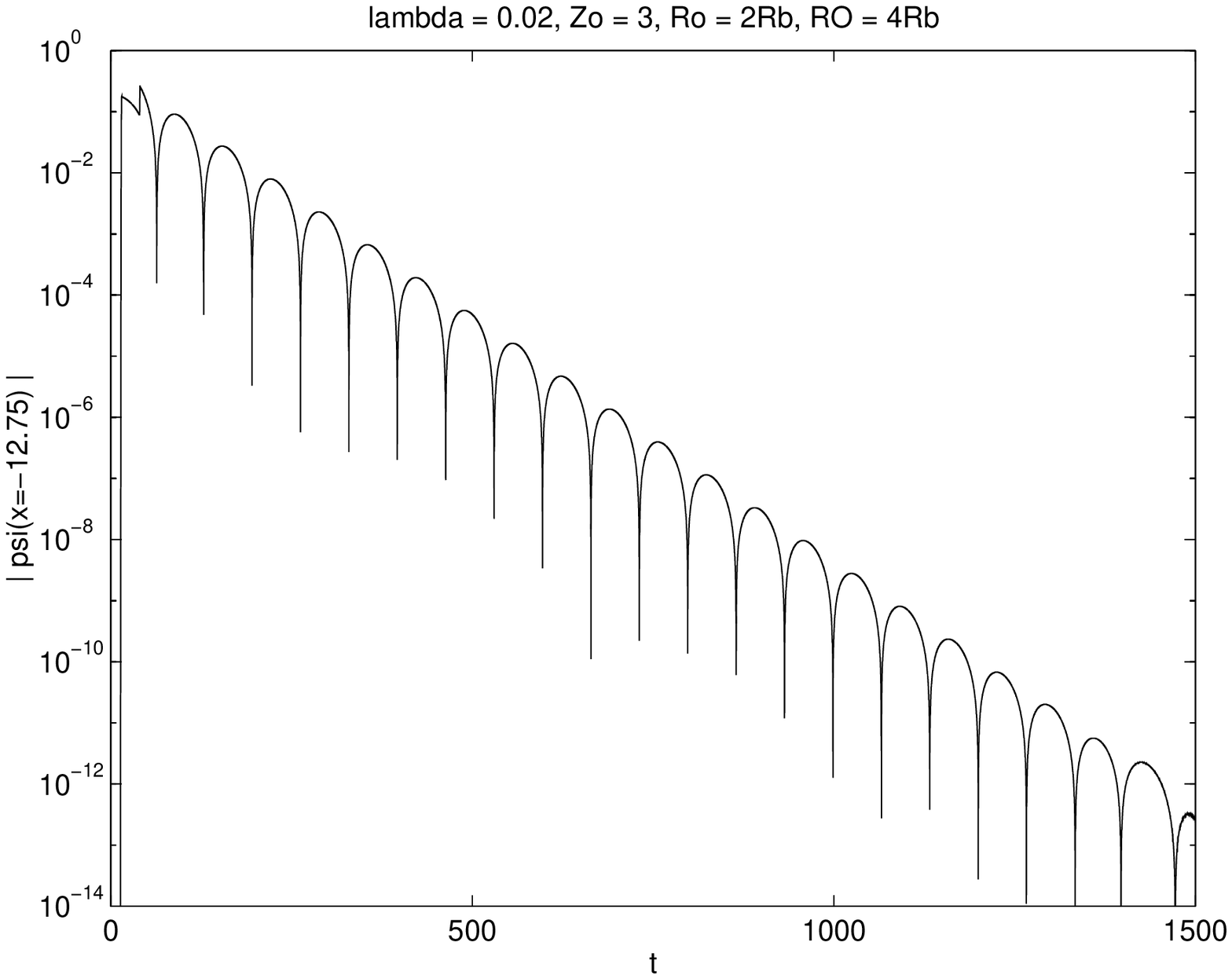,height=10cm} \hfill \mbox{}
    \caption{Exponential falloff using the Neumann condition}
    \label{Neumann1}
  \end{figure}

\begin{figure}[htbp]
    \hfill \psfig{file=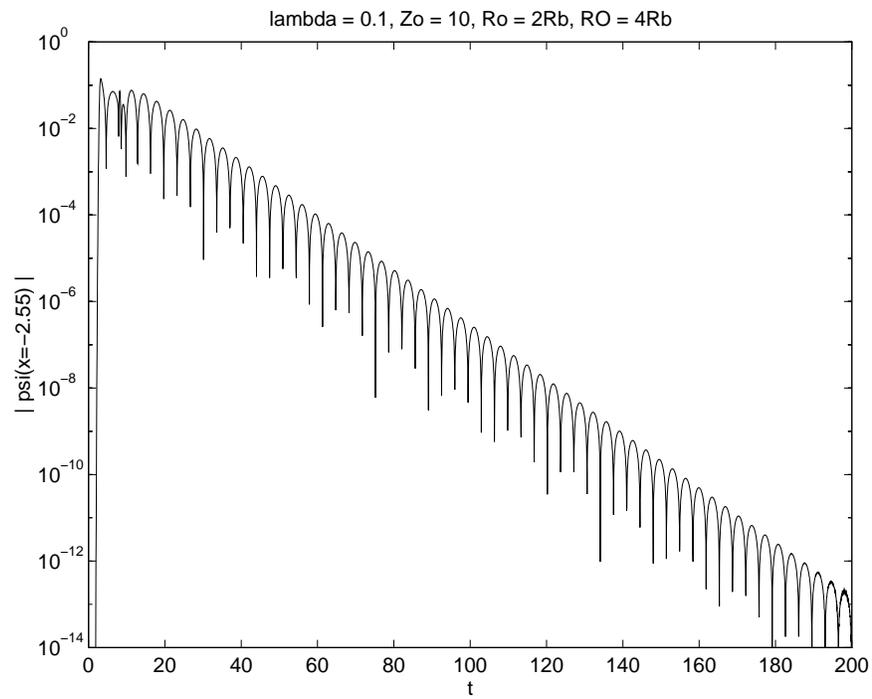,height=10cm} \hfill \mbox{}
    \caption{The same as in figure \ref{Neumann1} but with different
     parameters}
    \label{Neumann2}
  \end{figure}

  Figures \ref{Neumann1} and \ref{Neumann2} illustrate the
  numerical results using the approximate potential
  (\ref{Vapprox}) and the Neumann boundary condition at $x = 0$.
  It is clear from the two graphs that the scalar wave also
  exponentially decays when the Neumann condition is used.

  \subsection{Higher order correction}
 
    The results of the previous subsection are based upon the use of
    the potential barrier $V(x) = V_0\,\exp(2\,\lambda\,x)$ which
    is an approximation to 
    \begin{eqnarray}
      V(x) & = &
      V_0\,\frac{\exp(2\,\lambda\,x)}{\RdBk{1+\exp(2\,\lambda\,x)}^2} \period
    \end{eqnarray}
    Since $\lambda > 0$ but $x < 0$, $\exp(2\,\lambda\,x)$ is
    always less than one. Therefore $V(x)$ has a converging series
    expansion
    \begin{eqnarray}
      V(x) & = &
      V_0\,\SqBk{\exp(2\,\lambda\,x)-2\,\exp(4\,\lambda\,x)+3\,\exp(6\,\lambda\,x)-4\,\exp(8\,\lambda\,x)+O\!\RdBk{\exp(10\,\lambda\,x)}} \period
    \end{eqnarray}
    As a result we may iteratively solve the wave equation,
    treating $V_0\,\exp(2\,\lambda\,x)$ as the lowest order term,
    $\minus 2\,V_0\,\exp(4\,\lambda\,x)$ as the first correction
    to this approximation, and so on.

    For clarity, we define
    \begin{eqnarray}
      V_0(x) & \equiv & V_0\,\exp(2\,\lambda\,x) \comma \\
      V_1(x) & \equiv & \minus 2\,V_0\,\exp(4\,\lambda\,x) \comma \\
      \tilde{D} & \equiv & \minus \omega^2 - \di_{xx} + V_0(x) + V_1(x) \comma\\
      \tilde{D}_0 & \equiv & \minus \omega^2 - \di_{xx} + V_0(x) \period
    \end{eqnarray}
    Therefore the equation $\tilde{D} f(x;\omega) = 0$ has a
    representation
    \begin{eqnarray}
      f(x;\omega) & = &
      f_0(x;\omega)
      + \int_{\minus \infty}^x \frac{U_+(x)\,U_{\minus}(\xi)-U_{\minus}(x)\,U_+(\xi)}{W(\omega;U_{\minus},U_+)}\,V_1(\xi)\,f(\xi;\omega)\,d\xi
    \end{eqnarray}
    at $x = \minus \infty$. The functions $U_+(x)$ and $U_{\minus}(x)$
    satisfy the equation $\tilde{D}_0 U_+(x) = \tilde{D}_0 U_{\minus}(x) = 0$.
    The other function $f_0(x;\omega)$ satisfies the same equation
    as $U_+(x)$ and $U_{\minus}(x)$ and also satisfies the
    boundary conditions at $x \rightarrow \minus \infty$. In other
    words, $f_0(x;\omega)$ is the solution (\ref{BTZf0}) in our
    case. Similarly the solution around $x = 0$ is
    \begin{eqnarray}
      g(x;\omega) & = &
      g_0(x;\omega)
      - \int_x^0 \frac{U_+(x)\,U_{\minus}(\xi)-U_{\minus}(x)\,U_+(\xi)}{W(\omega;U_{\minus},U_+)}\,V_1(\xi)\,g(\xi;\omega)\,d\xi \comma
    \end{eqnarray}
    where $g_0(x;\omega)$ is just (\ref{BTZg0}). We first compute
    the Wronskian $W(\omega;g,f)$ for $f(x;\omega)$ and
    $g(x;\omega)$ above. Since this Wronskian is $x$-independent,
    the simplest way to compute it is to evaluate the quantity at
    the point $x = 0$. It is not difficult to show that
    \begin{eqnarray}
      W(\omega;g,f) & = &
      W(\omega;g_0,f_0)
      + \left. W(\omega;g_0,f)(x) \right|_{x = \minus \infty}^{x = 0}
      \; = \;
      W(\omega;g_0,f_0)
      - \left. W(\omega;g,f_0)(x) \right|_{x = \minus \infty}^{x = 0}
      \period
    \end{eqnarray}
    Since $g(x;\omega)$ and $f(x;\omega)$ satisfy a differential
    equation which differs from that satisfied by $g_0(x;\omega)$ and
    $f_0(x;\omega)$, the Wronskians $W(\omega;g_0,f)$ and
    $W(\omega;g,f_0)$ are functions of $x$ in general.
    Therefore the correction $V_1(x)$ in the potential barrier
    induces an extra ($x$-independent) term in the 
    Wronskian $W(\omega;g,f)$.

    The first Born approximations for $f(x;\omega)$ and
    $g(x;\omega)$ read
    \begin{eqnarray}
      f(x;\omega) & \approx &
      f_1(x;\omega) \; = \;
      f_0(x;\omega)
      + \frac{2\,\lambda^2}{V_0}\,\int_0^{z(x)} \SqBk{K_\nu\!\RdBk{z(x)}\,I_\nu(s)-I_\nu\!\RdBk{z(x)}\,K_\nu(s)}\,\hat{f_0}(s;\omega)\,s^3\,ds
      \label{BTZf1}
    \end{eqnarray}
    and
    \begin{eqnarray}
      g(x;\omega) & \approx &
      g_1(x;\omega) \; = \;
      g_0(x;\omega)
      - \frac{2\,\lambda^2}{V_0}\,\int_{z(x)}^{Z_0} \SqBk{K_\nu\!\RdBk{z(x)}\,I_\nu(s)-I_\nu\!\RdBk{z(x)}\,K_\nu(s)}\,\hat{g_0}(s;\omega)\,s^3\,ds
      \label{BTZg1}
    \end{eqnarray}
    if we choose $U_{\minus}(x) = I_\nu\!\RdBk{z(x)}$ and
    $U_+(x) = K_\nu\!\RdBk{z(x)}$. The functions
    $\hat{f_0}(s;\omega)$ and $\hat{g_0}(s;\omega)$ are defined
    via the equations
    \begin{eqnarray}
      \hat{f_0}\!\RdBk{z(x);\omega} & = & f_0(x;\omega) \; = \;
      {\cal N}_f\,I_\nu\!\RdBk{z(x)} \comma \\
      \hat{g_0}\!\RdBk{z(x);\omega} & = & g_0(x;\omega) \; = \;
      {\cal N}_g\,\SqBk{I_\nu(Z_0)\,K_\nu\!\RdBk{z(x)} - K_\nu(Z_0)\,I_\nu\!\RdBk {z(x)}} \period
    \end{eqnarray}
    As a result, the Wronskian $W(\omega;g,f)$ can be approximated by
    \begin{eqnarray}
      W(\omega;g,f) \; \approx \;
      W(\omega;g_0,f_0)
      + \left. W(\omega;g_0,f_1)(x) \right|_{\minus \infty}^0
      \; = \;
      W(\omega;g_0,f_0)
      + \int_{\minus \infty}^0 g_0(x;\omega)\,f_0(x;\omega)\,V_1(x)\,dx \period
    \end{eqnarray}
    This Wronskian has roots on the lower half $\omega$-plane.
    Thus there is quasi-normal ringing effect in the wave tail in
    this 3D black hole configuration with the approximate
    potential function  $V(x)\approx V_0(x) + V_1(x)$. 
    Moreover, since $g_1$ and $f_1$ have no cut, the
    falloff rate of the conformal wave is also exponential.

\begin{figure}[htbp]
      \hfill \psfig{file=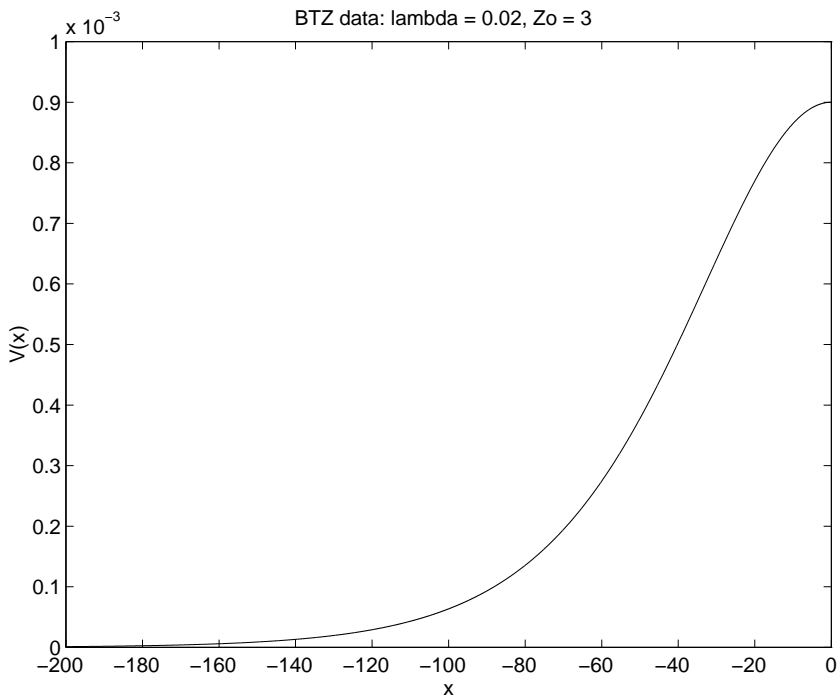,height=10cm} \hfill \mbox{}
      \caption{The shape of the exact 3D black hole potential $V(x)$}
      \label{BTZ1}
    \end{figure}

\begin{figure}[htbp]
      \hfill \psfig{file=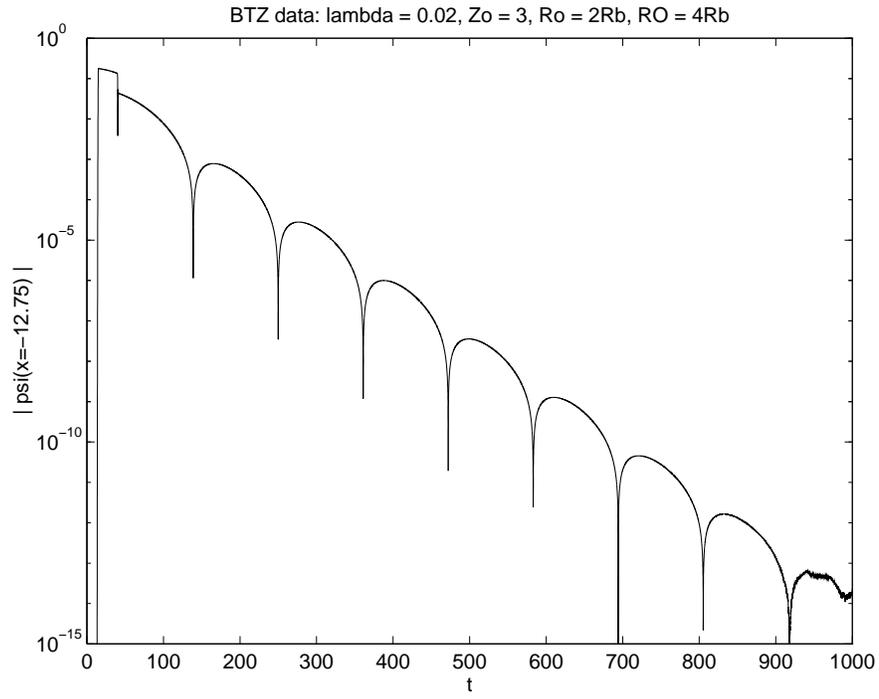,height=10cm} \hfill \mbox{}
      \caption{Exponential decay of the scalar wave $\psi$
	       in the static 3D black hole background}
      \label{BTZ2}
    \end{figure}

\begin{figure}[htbp]
      \hfill \psfig{file=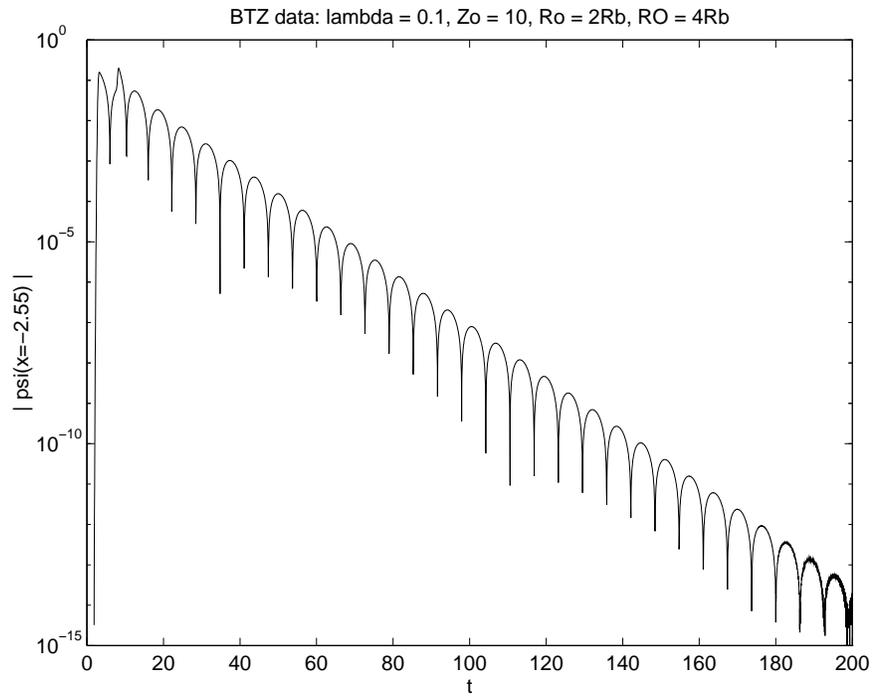,height=10cm} \hfill \mbox{}
      \caption{Scalar wave decay as in fig \ref{BTZ2}
	       but with different parameters}
      \label{BTZ3}
    \end{figure}

    The shape of the exact potential function $V(x)$ is given on
    figure \ref{BTZ1} with parameters $\lambda = 0.02$ and
    $Z_0 = 3$. Figures \ref{BTZ2} and \ref{BTZ3} are results of
    the numerical calculations using the exact 3D black hole
    potential (\ref{BTZpotential}). The methodology underlying
    these computations is identical to the previous case that uses
    the approximate exponential potential function. The parameters
    and boundary conditions used in this case are also the same 
    as the cases in figures \ref{p_BTZ1} and \ref{p_BTZ2}. 
    Once again the numerical calculations show the exponential falloff 
    of the scalar wave in the static 3D background.


  \subsection{Spinning 3D Black Hole Background}
    When $J \neq 0$, i.e. the black hole rotates, we also find a
    late time exponential decay rate, as we shall now demonstrate.
    
    If we assume that
    \begin{eqnarray}
      \Psi(t,r,\phi) & = & \frac{\psi(t,r)}{\sqrt{r}} \comma
    \end{eqnarray}
    that is to say there is no ``spherical harmonic" component to
    the wave, the conformal scalar wave equation
    $\Del^2 \Psi = R\,\Psi/8$ will reduce to (\ref{WE5}). The
    tortoise coordinate $x$ is defined as before but the potential
    barrier becomes
    \begin{eqnarray}
      V_e\!\RdBk{x(r)} & = &
      \minus \frac{3\,|\Lambda|}{4} + \frac{1}{2\,r}\,\di_r N(r) - \frac{N(r)}{4\,r^2} \period
    \end{eqnarray}

    In this spinning case, the black hole has two horizons $R_\pm$
    which are given by the equation
    \begin{eqnarray}
      {R_\pm}^2 & = &
      \frac{1}{2\,|\Lambda|}\,\SqBk{M \pm \sqrt{M^2-|\Lambda|\,J^2}} \period
    \end{eqnarray}
    The lapse function $N(r)$ can then be written in terms of
    $R_+$ and $R_{\minus}$ as
    \begin{eqnarray}
      N(r) & = &
      \frac{|\Lambda|}{r^2}\,\RdBk{r^2-{R_+}^2}\,\RdBk{r^2-{R_{\minus}}^2}
      \period
    \end{eqnarray}
    As a result, the tortoise coordinate reads
    \begin{eqnarray}
      x(r) & = &
      \frac{1}{2\,|\Lambda|\,\RdBk{{R_+}^2-{R_{\minus}}^2}}\,\SqBk{R_+\,\ln\!\RdBk{\frac{r-R_+}{r+R_+}}-R_{\minus}\,\ln\!\RdBk{\frac{r-R_{\minus}}{r+R_{\minus}}}}
    \end{eqnarray}
    which has an inverse of the form
    \begin{eqnarray}
      \frac{R_+}{r(x)} & = &
      \frac{Y}{1-\sigma^2}\,\sum_{n=0}^\infty a_n(\sigma)\,Y^{2\,n} \comma \\
      \sigma & = & \frac{R_{\minus}}{R_+} \; < \; 1 \comma \\
      Y & = & \frac{1-\exp(2\,\lambda\,x)}{1+\exp(2\,\lambda\,x)}
      \; < \; 1 \comma \\
      \lambda & = & |\Lambda|\,R_+\,(1-\sigma^2) \; > \; 0 \period
    \end{eqnarray}
    The coefficients $a_n(\sigma)$ which read
    \begin{eqnarray}
      a_0(\sigma) \; = \; 1
      \comma & \hspace{1cm} &
      a_1(\sigma) \; = \; \minus \sigma^2\,\frac{3-\sigma^2}{3\,(1-\sigma^2)^2}
      \comma \\
      a_2(\sigma) \; = \;
      \sigma^4\,\frac{25-17\,\sigma^2+3\,\sigma^4}{15\,(1-\sigma^2)^4}
      \comma & \hspace{1cm} &
      a_3(\sigma) \; = \;
      \minus \sigma^6\,\frac{1008-1039\,\sigma^2+368\,\sigma^4-45\,\sigma^6}{315\,(1-\sigma^2)^6}
      \comma  \hspace{5mm} \cdots
    \end{eqnarray}
    are of order $O\!\RdBk{\sigma^{2\,n}}$. If $\sigma$ is small
    enough, we can employ the approximation
    \begin{eqnarray}
      \frac{1}{r(x)} & \approx & \frac{Y}{R_+} \period
    \end{eqnarray}
    If we substitute this approximation into the potential barrier
    $V(x)$, we will obtain
    \begin{eqnarray}
      V(x) & \approx &
      \Lambda^2\,\frac{({R_+}^2-{R_{\minus}}^2)\,({R_+}^2-4\,{R_{\minus}}^2)}{{R_+}^2}\,\frac{\exp(2\,\lambda\,x)}{\RdBk{1+\exp(2\,\lambda\,x)}^2} \nonumber \\&&
      + \; 12\,\Lambda^2\,(2\,{R_+}^2-3\,{R_{\minus}}^2)\,\frac{{R_{\minus}}^2}{{R_+}^2}\,\frac{\exp(4\,\lambda\,x)}{\RdBk{1+\exp(2\,\lambda\,x)}^4}
      + 80\,\Lambda^2\,\frac{{R_{\minus}}^4}{{R_+}^2}\,\frac{\exp(6\,\lambda\,x)}{\RdBk{1+\exp(2\,\lambda\,x)}^6} \period
    \end{eqnarray}
    It is obvious that the first approximation of this potential has a
    form $V(x) \approx V_0\,\exp(2\,\lambda\,x)$ while the second one is
    $V(x) \approx V_0\,\exp(2\,\lambda\,x) + V_1\,\exp(4\,\lambda\,x)$.
    According to our previous analysis, we may conclude that the
    decay rate at late time is also exponential rather than a
    power-law.

    \begin{figure}[htbp]
      \hfill \psfig{file=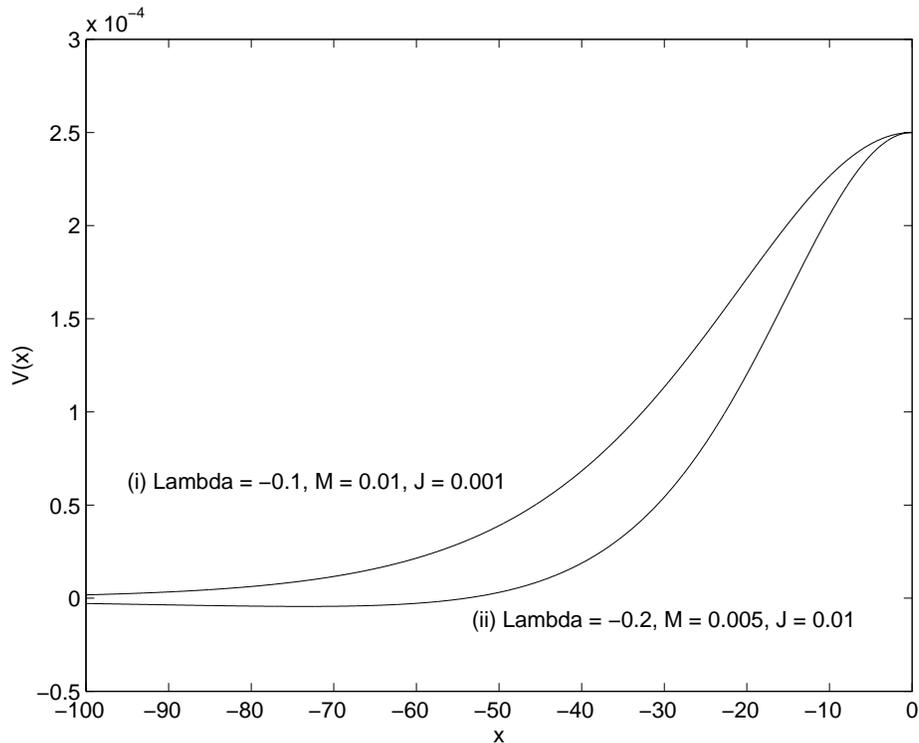,height=10cm} \hfill \mbox{}
      \caption{Potential function $V(x)$ of two spinning
	       3D black hole backgrounds with
	       (i) $|\Lambda|\,J^2 / M^2 = 10^{\minus 3}$ and
	       (ii) $|\Lambda|\,J^2 / M^2 = 0.8$}
      \label{SBTZ1}
    \end{figure}

\begin{figure}[htbp]
      \hfill \psfig{file=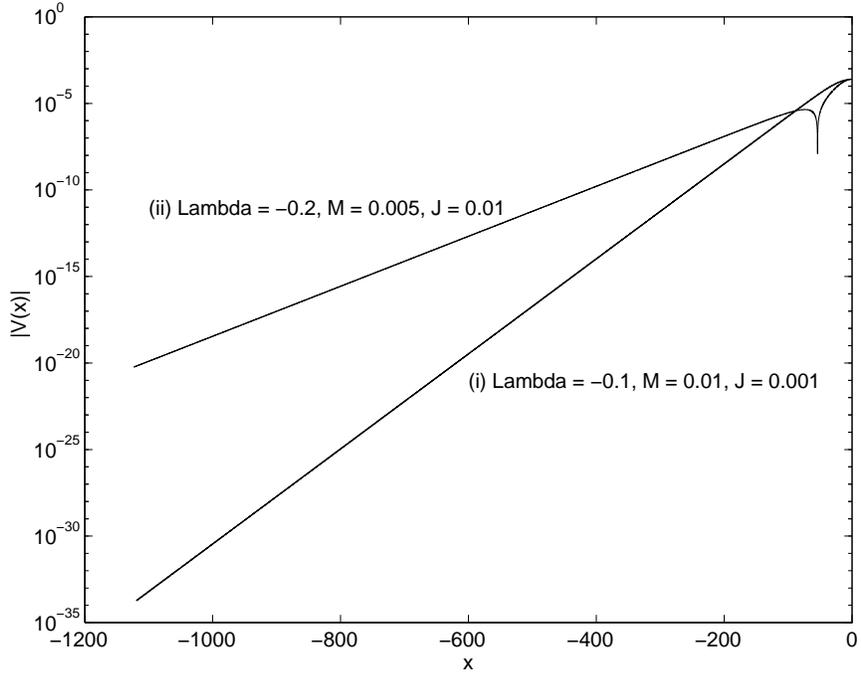,height=10cm} \hfill \mbox{}
      \caption{Decaying exponential behaviour of $|V(x)|$ 
               near the black hole event horizon}
      \label{SBTZ2}
    \end{figure}

\begin{figure}[htbp]
      \hfill \psfig{file=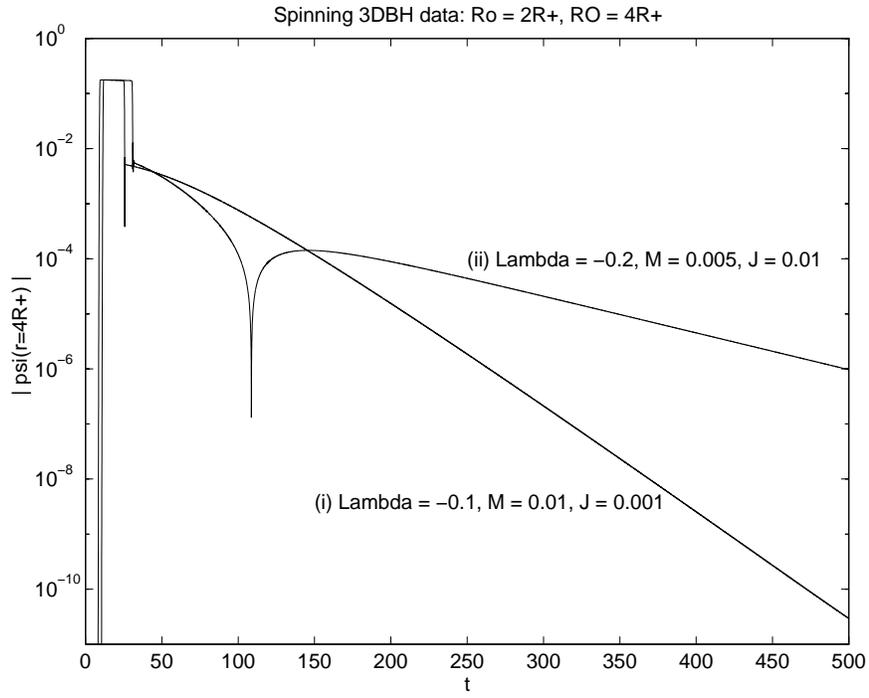,height=10cm} \hfill \mbox{}
      \caption{Exponential decay of conformal scalar waves
	       in spinning 3D black hole geometry}
      \label{SBTZ3}
    \end{figure}

    Figure \ref{SBTZ1} is the graph of the potential functions of
    this spinning case with zero angular harmonic. One can show
    that when the ratio $|\Lambda|\,J^2/M^2 > 16/25$, $V_e$
    becomes negative for some $r>R_+$. This behaviour is 
    illustrated in figure \ref{SBTZ1}, where 
    $V(x)$ becomes negative when $|x|$ is sufficiently large.
    In figure \ref{SBTZ2}, we can see that the
    potential functions in both cases vanish at an exponential
    rate towards the event horizon. As usual, we put a Gaussian
    impulse at a distance twice of the outer horizon $R_+$ and the
    numerical response of the scalar waves over time at a distance
    $4\,R_+$ is shown in figure \ref{SBTZ3}. This figure shows an
    exponential decay of the wave although its appearance differs
    from the previous graphs because there are no angular
    harmonics (i.e. $l = 0$). Indeed if we set $l = 0$ in the
    static 3D black hole case, we find that the falloff also looks
    like that in figure \ref{SBTZ3} because the pole that is
    closest to the real $\omega$-axis for the Green's function in
    frequency domain has no real part.

\section{Schwarzschild-Anti-de Sitter Background}
  Since the 3D black hole spacetime is asymptotically
  anti-de Sitter, one might expect that the late time falloff
  behaviour of any scalar wave in Schwarzschild-anti-de Sitter
  (SAdS) background has similar behaviour. In fact, the situation
  is quite different from the 3D case, in part because of the
  different dimensionality and in part because it is not possible
  to solve the wave equation (\ref{WE5}) exactly in this
  background. Although the potential function $V_e(r)$ in this
  case reads
  \begin{eqnarray}
    V_e(r) & = &
    \frac{2\,|\Lambda|}{3}\,(1-6\,\xi) + \frac{l\,(l+1)}{r^2}
    + \frac{2\,M}{r^3} \label{SAdSVe}
  \end{eqnarray}
  which is quite similar to (\ref{BTZVe}) in the 3D case, we
  cannot write $r$ as a function of $x$ in a closed form. That is
  to say given the lapse function
  \begin{eqnarray}
    N(r) \; = \;
    \frac{|\Lambda|}{3}\,r^2 + 1 - \frac{2\,M}{r}
    \; = \;
    \frac{|\Lambda|}{3}\,r^2 + 1 - \frac{R_b\,(3+|\Lambda|\,{R_b}^2)}{3\,r}
    \comma \label{SAdSmetric}
  \end{eqnarray}
  where $\Lambda < 0$ and the black hole radius $R_b$ satisfies
  the equation $N(R_b) = 0$, the tortoise coordinate
  \begin{eqnarray}
    x & = &
    \frac{R_b}{2\,(1+|\Lambda|\,{R_b}^2)}\,\ln\!\left|\,\frac{(r-R_b)^2}{r^2+R_b\,r+{R_b}^2+3/|\Lambda|}\,\right| \nonumber \\ &&
    + \; \frac{\sqrt{3}\,(2+|\Lambda|\,{R_b}^2)}{\sqrt{|\Lambda|}\,(1+|\Lambda|\,{R_b}^2)\,\sqrt{4+|\Lambda|\,{R_b}^2}}\,\SqBk{\arctan\!\RdBk{\frac{\sqrt{|\Lambda|}\,(2\,r+R_b)}{\sqrt{3}\,\sqrt{4+|\Lambda|\,{R_b}^2}}} - \frac{\pi}{2}}
  \end{eqnarray}
  has no closed form inverse, and so we are unable to write down
  an explicit expression for the potential $V(x)$.
  
  For the remainder of this section we set $\xi=1/6$ (i.e. we
  consider a conformal scalar field propagating on an SAdS background).

\begin{figure}[htbp]
    \hfill \psfig{file=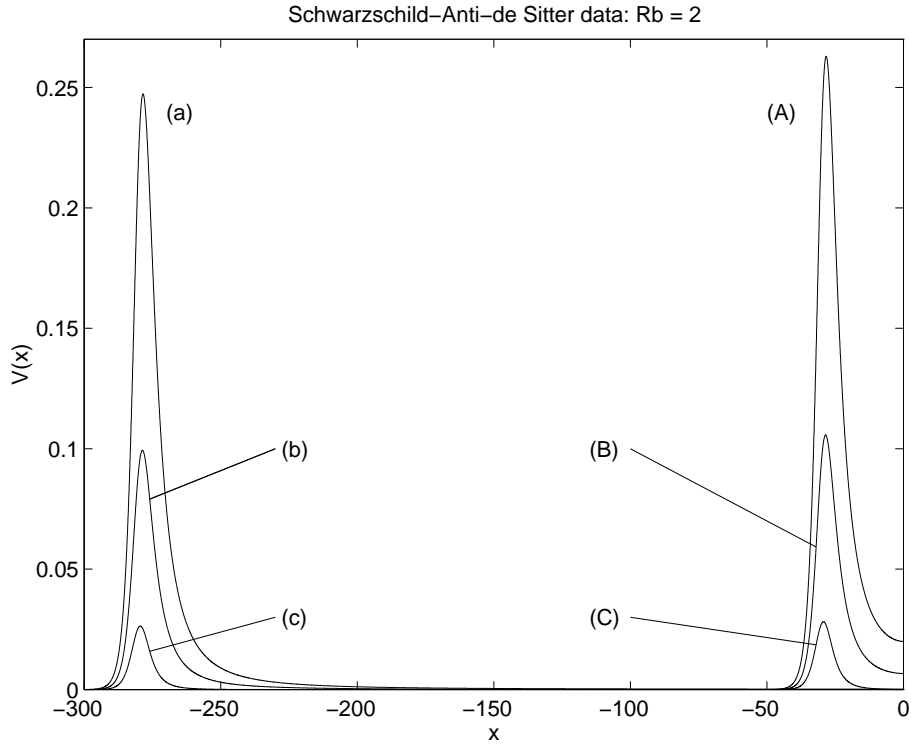,height=10cm} \hfill \mbox{}
    \caption{Potential functions $V(x)$ for the SAdS background.
		 The six potentials are generated with the parameters
		 (a) $\Lambda = \minus 10^{\minus 4}, l = 2$,
		 (b) $\Lambda = \minus 10^{\minus 4}, l = 1$,
		 (c) $\Lambda = \minus 10^{\minus 4}, l = 0$,
		 (A) $\Lambda = \minus 10^{\minus 2}, l = 2$,
		 (B) $\Lambda = \minus 10^{\minus 2}, l = 1$,
		 (C) $\Lambda = \minus 10^{\minus 2}, l = 0$.}
    \label{SAdS1}
  \end{figure}

  The most important difference between the 3D black hole
  background and SAdS background is the shape of the potential
  function $V(x)$. In both cases the potential functions are
  decreasing in an exponential manner toward the horizon. In the
  3D case (either spinning or static) this function attains a
  maximum at a distance $r = \infty$ ($x = 0$). When the
  background is SAdS, this is no longer true because spatial
  infinity (which is still given by $x=0$) is not the place at
  which $V(x)$ has an absolute maximum (for $x \in \Re^{\minus}$).
  As with the Schwarzschild black hole (figure \ref{Schwz2}),
  the potential function $V(x)$ attains a maximum not far away
  from the event horizon. In the SAdS case, the shape of the
  potential function $V(x)$ is given in figure \ref{SAdS1}.
  Unlike the Schwarzschild case, the tortoise coordinate $x$
  for the SAdS background is bounded above. Eventually all
  the outgoing waves that leave the black hole region will
  return towards it due to the boundary condition at $x = 0$.
  The returning wave will then reflect off of the potential
  barrier back toward spatial infinity. This is completely
  different behaviour from the 3D case, in which the incoming
  wave from spatial infinity $x = 0$ continues its journey
  to the black hole unhindered.

  We can see from figure \ref{SAdS1} that when $|\Lambda|$ is
  small (barriers (a)--(c)), the barrier maximum moves to the
  left, lengthening the travelling time from this maximum
  to spatial infinity. When $l$ vanishes (case (c) and (C)),
  $V(0) = 0$ because $V(0) = l\,(l+1)\,|\Lambda| / 3$ in general.
  Therefore barriers (a) and (b) have $V(0) \neq 0$ although this
  feature is not apparent on the graph due to the small size of
  $|\Lambda|$. The barrier height is considerably higher than
  the magnitude of $V(x)$ at $x = 0$, and this feature becomes
  more pronounced for large $l$. This causes the scalar wave to
  bounce back and forth in the region outside the barrier.
  However part of the scalar wave can surmount the barrier
  (thereby going into the black hole) because the barrier
  height is still finite. However it takes a long time for a
  significant portion of the wave to enter the black hole. 

\begin{figure}[htbp]
    \hfill \psfig{file=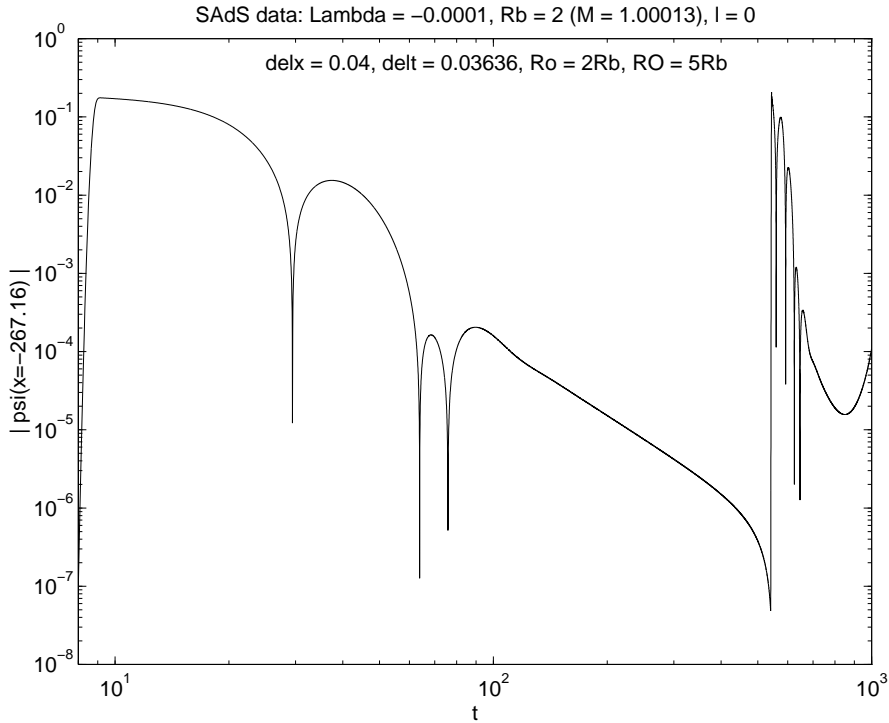,height=10cm} \hfill \mbox{}
    \caption{Scalar wave of $l = 0$ decays away in SAdS background
		 using Dirichlet condition at $x = 0$}
    \label{SAdS2}
  \end{figure}

\begin{figure}[htbp]
    \hfill \psfig{file=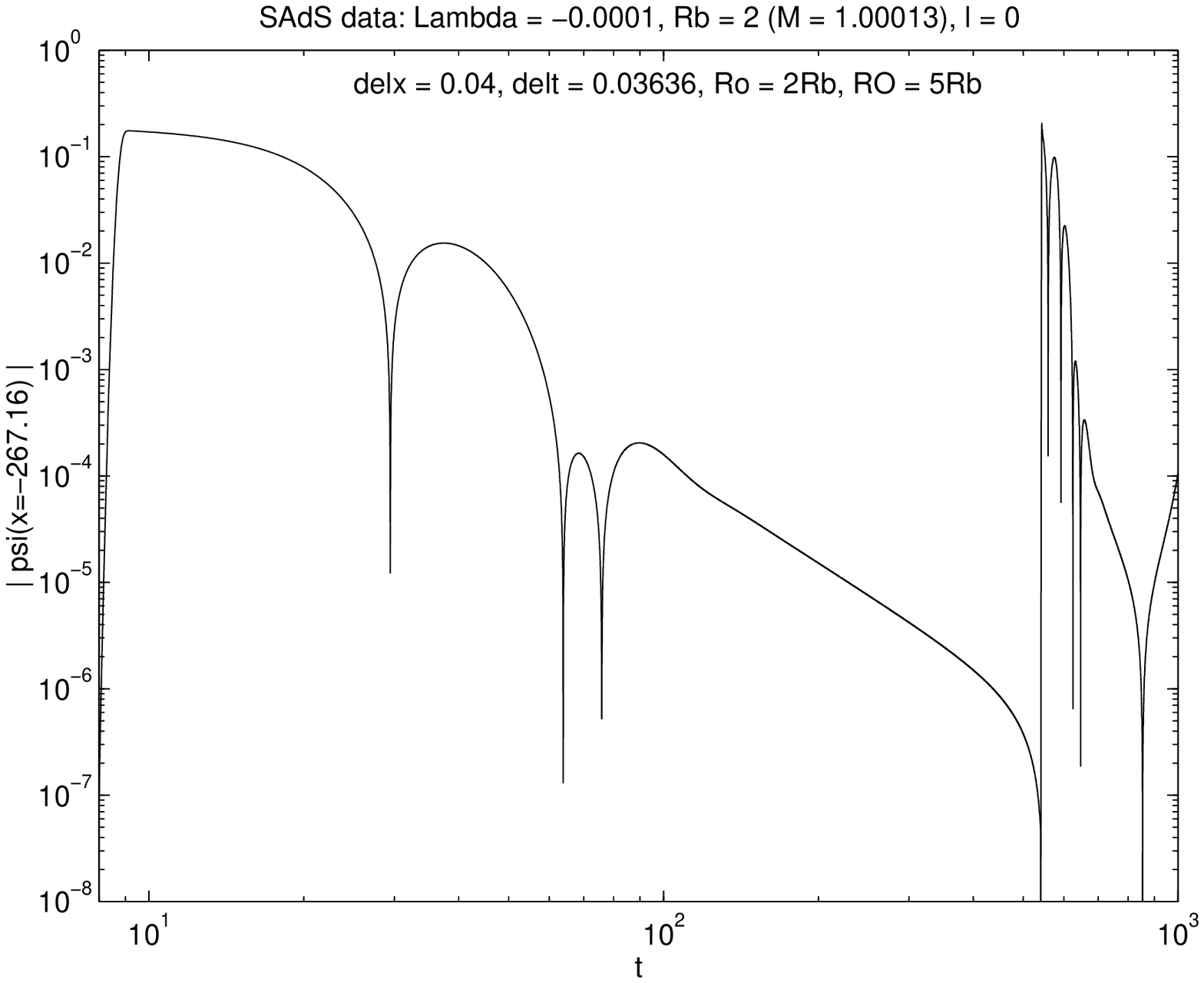,height=10cm} \hfill \mbox{}
    \caption{Scalar wave falloff pattern of $l = 0$ using
		 Neumann condition spatial infinity}
    \label{SAdS3}
  \end{figure}

  We solve the wave equation in this SAdS background numerically.
  The results of the numerical integration of equation (\ref{WE5})
  using (\ref{SAdSVe}) and (\ref{SAdSmetric}) are given in the
  next few graphs. Figures \ref{SAdS2} and \ref{SAdS3} show the
  falloff behaviour of a conformal scalar wave with $l = 0$
  initially located at a distance $r = 2\,R_b$. The computation
  for figure \ref{SAdS2} uses the Dirichlet condition at $x = 0$
  but Neumann boundary condition is employed for figure \ref{SAdS3}.
  Since the cosmological constant $|\Lambda|$ was chosen to be
  relatively small in both cases, namely $|\Lambda| = 10^{\minus 4}$,
  we can see on both graphs that there is clearly an inverse
  power-decay behaviour. According to the graph, this power-decay
  rate is roughly $t^{\minus 3}$ which agrees with the one in
  Schwarzschild case. However this inverse power-decay does not
  last very long after the return of the outgoing wave from
  spatial infinity. Both diagrams show this returning wavefront.

\begin{figure}[htbp]
    \hfill \psfig{file=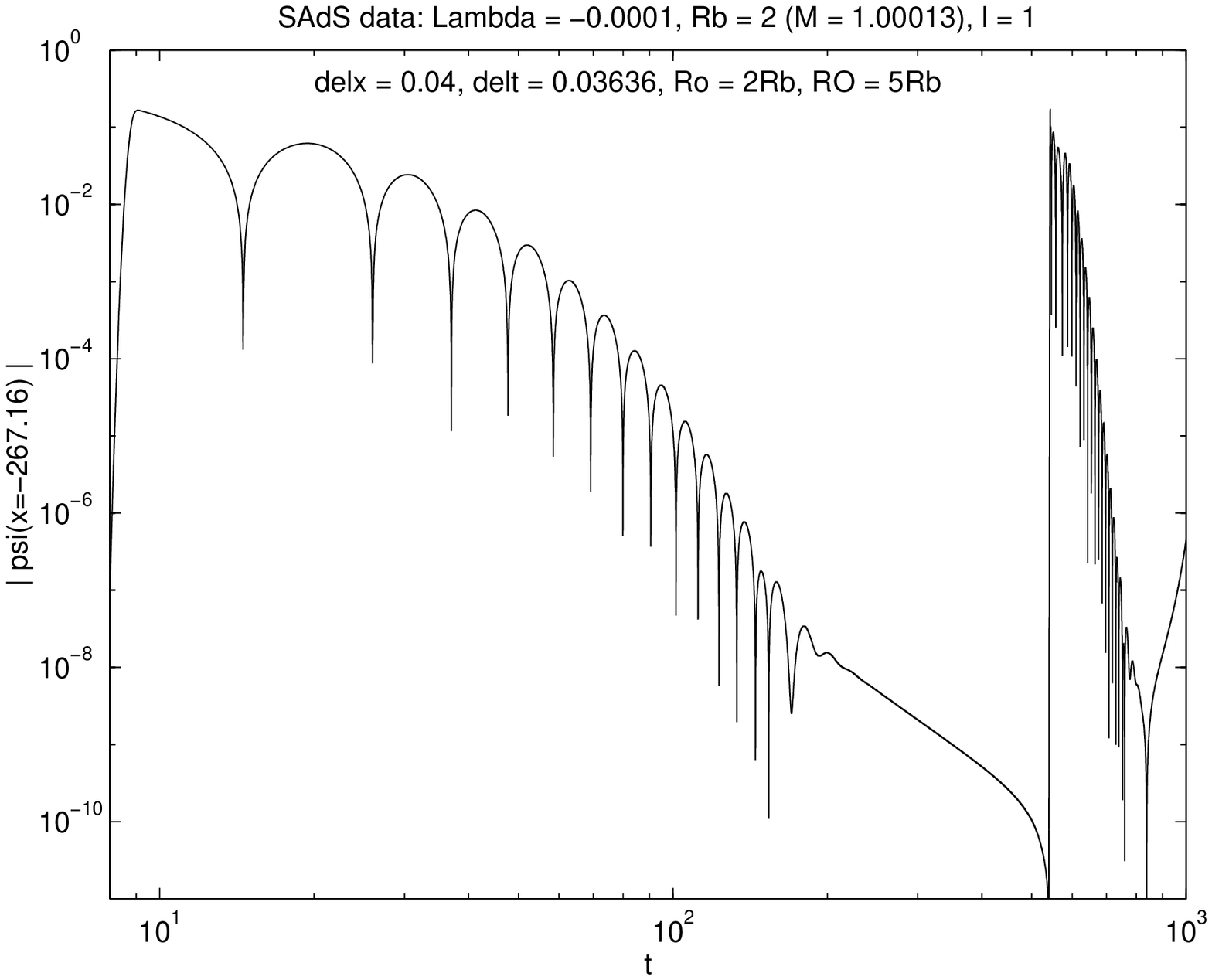,height=10cm} \hfill \mbox{}
    \caption{Loglog graph of the decay behaviour in SAdS background
		 using $l = 1$ and Dirichlet condition}
    \label{SAdS4}
  \end{figure}

\begin{figure}[htbp]
    \hfill \psfig{file=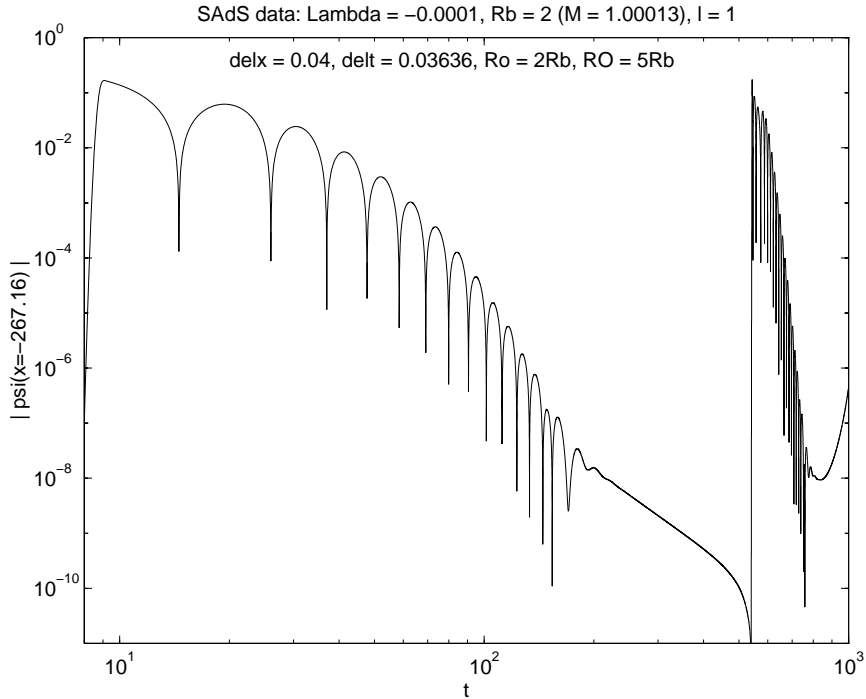,height=10cm} \hfill \mbox{}
    \caption{Similar graph to Figure \ref{SAdS4} but using
		 Neumann condition}
    \label{SAdS5}
  \end{figure}

  For small $|\Lambda|$ and nonzero $l$, the falloff behaviour
  resembles the case of $l = 0$. Initially there is a ringing
  effect (due to the quasi-normal modes) followed by inverse
  power-decay behaviour as shown in figures \ref{SAdS4} and
  \ref{SAdS5}.

\begin{figure}[htbp]
    \hfill \psfig{file=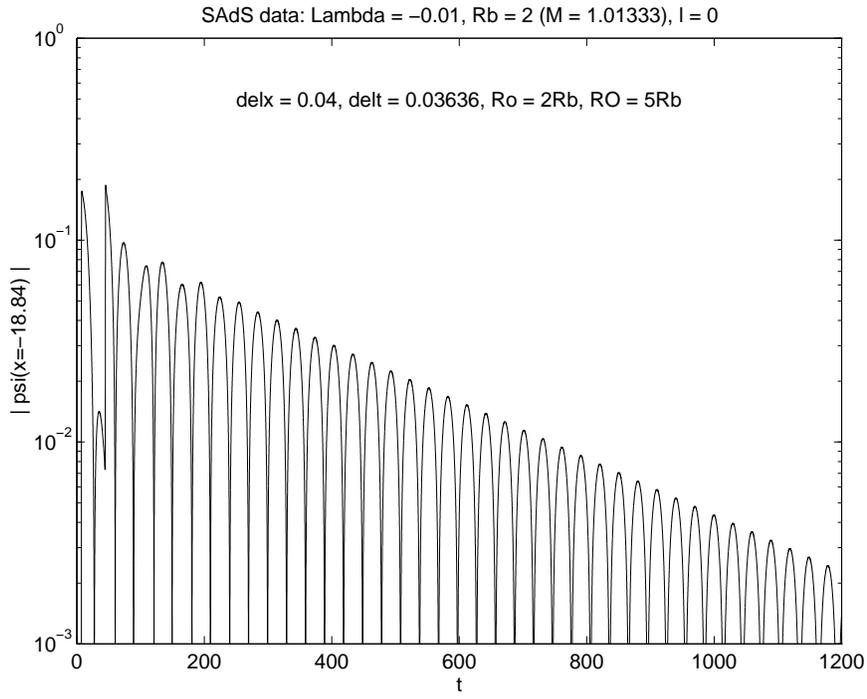,height=10cm} \hfill \mbox{}
    \caption{Semilog graph of the decay behaviour in SAdS background
             using $l = 0$ and Dirichlet condition}
    \label{SAdS6}
  \end{figure}

\begin{figure}[htbp]
    \hfill \psfig{file=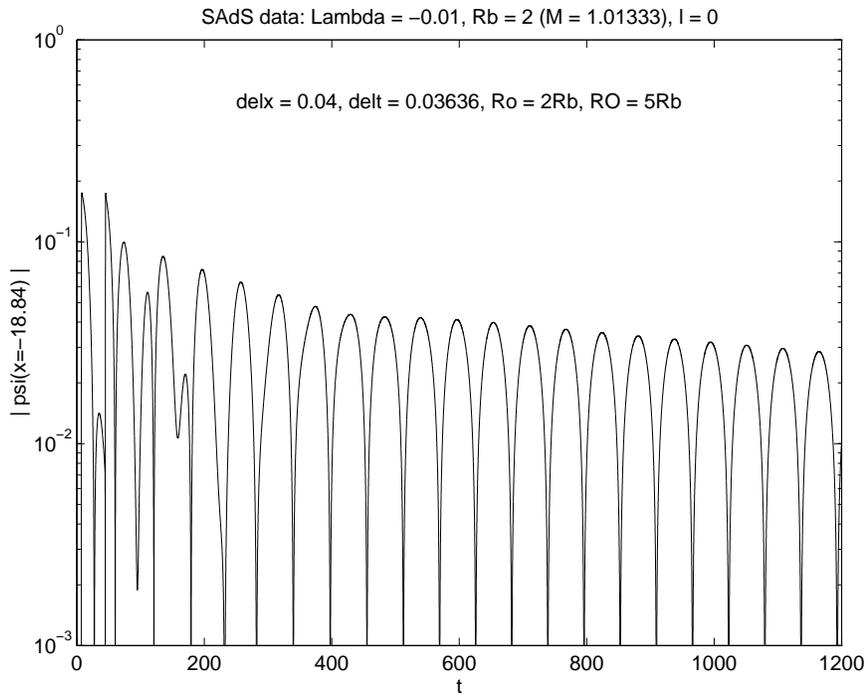,height=10cm} \hfill \mbox{}
    \caption{Conformal scalar wave decay behaviour
             using $l = 0$ and Neumann condition}
    \label{SAdS7}
  \end{figure}

\begin{figure}[htbp]
    \hfill \psfig{file=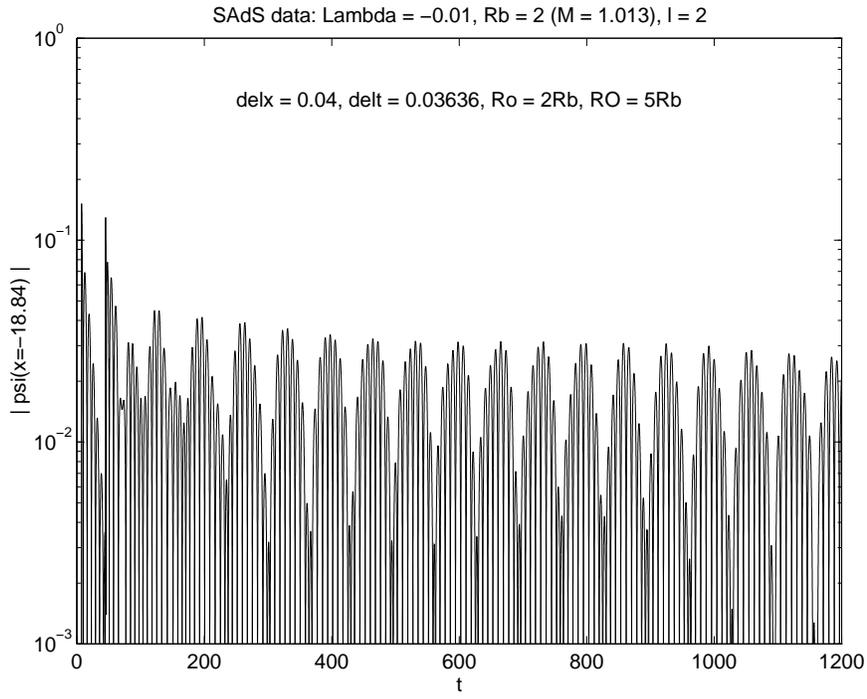,height=10cm} \hfill \mbox{}
    \caption{Semilog graph of the decay behaviour using
	     non-zero $l$ and Dirichlet condition}
    \label{SAdS8}
  \end{figure}

\begin{figure}[htbp]
    \hfill \psfig{file=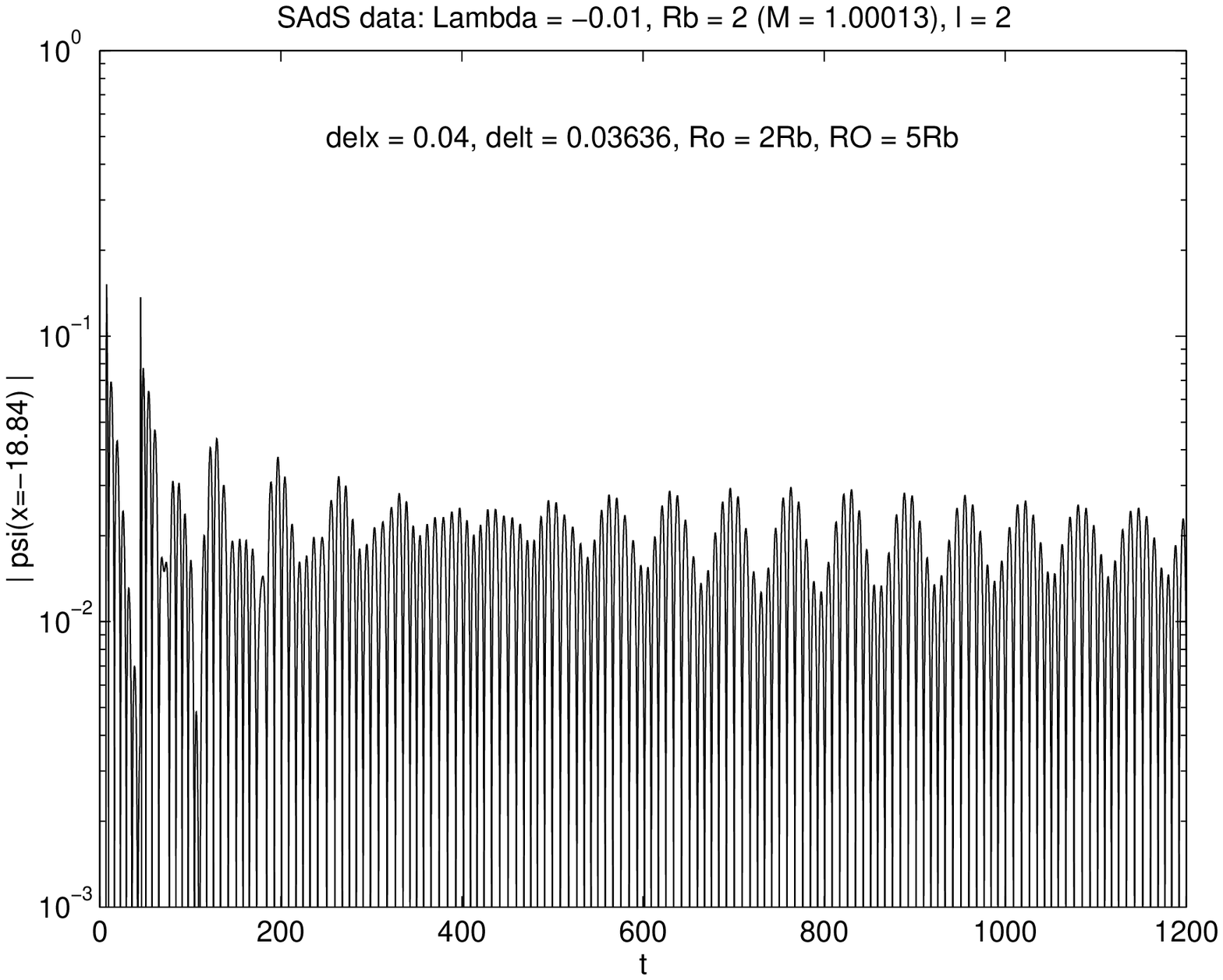,height=10cm} \hfill \mbox{}
    \caption{This semilog graph is the Neumann-analogue of Figure \ref{SAdS8}}
    \label{SAdS9}
  \end{figure}

  Since the finite height of the potential maximum allows the
  scalar wave to surmount the barrier and leave the trapped region
  which is the exterior outside the barrier maximum, we expect
  that the peak value of the second returning wave is smaller than
  that of the first wave. However it is unclear from simple
  inspection of figures \ref{SAdS2} to \ref{SAdS5} whether or not
  this is the case. Figures \ref{SAdS6} to \ref{SAdS9} are the
  numerical results we obtained using the potential functions (A)
  and (C) in figure \ref{SAdS1}. We compute the large $|\Lambda|$ case
  with both Dirichlet (figures \ref{SAdS6} and \ref{SAdS8})
  and Neumann (\ref{SAdS7} and \ref{SAdS9}) boundary conditions. 
  For this case there are more returning waves
  in a reasonable amount of CPU time. From these graphs, we can
  see that the scalar wave does indeed decrease but over a much larger
  time scale. Figures \ref{SAdS6} and \ref{SAdS7} indicate that
  the peak-height has an approximate exponential falloff for $l=0$.
  For $l>0$, the peak-height has a more complicated behaviour illustrated
  in figures \ref{SAdS8} and \ref{SAdS9}.  Over long time scales
  the maximum peak-height has a very mild approximate exponential
  falloff.


\section{Conclusion}

  The results of this work indicate clearly that the asymptotics
  of a given spacetime have a considerable influence on the late
  time falloff behaviour of a scalar wave. The early work of Price
  showing that waves in Schwarzschild background decay away
  according to an inverse power rate was recently extended by
  Ching {\sl et al.} who found that the inverse power-law can also
  be modified by a logarithmic term in some black hole
  configurations. We have been concerned in this paper with the
  falloff behaviour of conformal scalar waves in a background
  geometry which is asymptotically anti-de Sitter. We found that
  the waves in uncharged static and spinning 3D black hole
  backgrounds die out at an exponential rate. This conclusion is
  supported by both analytical and numerical computation. For
  conformal scalar waves in a Schwarzschild-anti-de Sitter
  background our numerical analysis demonstrates that the falloff
  pattern over short time scales obeys (after some ringing) a
  power-law falloff. However over longer time scales the outgoing
  wave returns from spatial infinity. The falloff of the peak amplitude
  in this case is neither inverse power nor exponential. However
  there is a very weak exponential decrease in the maximal 
  peak amplitude.   This decay pattern is so complicated that 
  further investigation will be
  needed to more precisely determine its dependence on $t$.

  Previous investigations of mass inflation in a charged spinless
  3D black hole \cite{Husain} and an uncharged spinning 3D black
  hole \cite{MI2} assumed a power-law falloff rate for the scalar
  wave. Under such assumptions these investigations demonstrated
  that mass inflation occurred in the corresponding geometries. We
  have shown here that the assumption of power-law falloff in the
  uncharged static and spinning 3D black hole background is not
  valid and must be replaced by an exponential falloff.

  For a rotating 3D black hole we have shown that a
  $\phi$-independent scalar field also has late time exponential
  decay. In this particular setup, one can recalculate the mass
  inflation rate using the correct late time falloff rate. It is
  not difficult to show that the new inflation rate reads
  $m(v) \propto \exp( (\kappa-\alpha)\,v )$, where $\kappa$ is the
  surface gravity at the Cauchy horizon of the black hole. The
  parameter $\alpha$ comes from the exponential falloff rate for
  the scalar wave (which reads $\exp(\minus \alpha\,t)$) and is
  determined (say in the $J = 0$ case) by setting $I_\nu(Z_0)=0$
  for the relevant $Z_0$ and finding (numerically) the value of
  $\nu$ with the maximum negative real part. For example when
  $\Lambda = \minus 0.1$, $M = 0.01$ and $J = 0.001$, the surface
  gravity at the Cauchy horizon is $1.999$ but the parameter
  $\alpha$ is only $0.04$ according to figure \ref{SBTZ3}.
  Table \ref{table1} shows some other values of $\alpha$ and
  $\kappa$ in the rotating 3D black hole backgrounds. All these
  cases are consistent to the claim that the surface gravity
  $\kappa$ is always greater than the exponential falloff rate
  $\alpha$. Therefore the attenuation from the falloff effect does
  not halt mass inflation and the basic conclusions in \cite{MI2}
  remain unchanged. Rather the mass inflation rate is slowed down
  relative to spacetimes with power-law falloff.
  \begin{table}
    \hfill \vbox{\tabskip=0pt \offinterlineskip
        \def\tablerule{\noalign{\hrule}} \halign to400pt{
                \strut #& \vrule # \tabskip=2mm plus2mm &
                \hfil # \hfil & \vrule # &
		\hfil # \hfil & \vrule # &
		\hfil # \hfil & \vrule # &
		\hfil # \hfil & \vrule # &
		\hfil # \hfil & \vrule # &
                \hfil # \hfil & \vrule # \tabskip=0pt \cr \tablerule
		&& $|\Lambda|$ && $M$ && $J$ && $|\Lambda|\,J^2 / M^2$ && $\alpha$ && $\kappa$ & \cr \tablerule
		&& 0.1 && 0.5 && 0.001 && $4 \times 10^{\minus 7}$ && 0.3171 && 707.107 & \cr \tablerule
		&& 0.1 && 0.1 && 0.001 && $10^{\minus 5}$ && 0.1413 && 63.2452 & \cr \tablerule
		&& 1.0 && 1.0 && 0.01 && $10^{\minus 4}$ && 1.417 && 199.987 & \cr \tablerule
		&& 0.1 && 0.01 && 0.001 && $10^{\minus 3}$ && 0.04422 && 1.99875 & \cr \tablerule
		&& 0.5 && 0.01 && 0.001 && $5 \times 10^{\minus 3}$ && 0.09852 && 1.99374 & \cr \tablerule
		&& 0.1 && 0.005 && 0.01 && 0.4 && 0.02034 && 0.0515936 & \cr \tablerule
		&& 0.2 && 0.005 && 0.01 && 0.8 && 0.01541 && 0.0268999 & \cr \tablerule
    }} \hfill \mbox{}
    \caption{Exponential falloff rate $\alpha$ which is found by
	     graphical method is always less than the surface
	     gravity $\kappa$ in rotating 3D black holes.}
    \label{table1}
  \end{table}

  We have numerically determined that $\alpha$ is an increasing
  function of $l$. Recall that for the approximate potential
  $V(x)$ in (\ref{Vapprox}), the dominant decay rate of the
  Green's function is determined by the root $\omega_0$ of the
  equation $I_{\minus i\,\omega / \lambda}(Z_0) = 0$ in the lower
  half $\omega$-plane that is closest to the real axis. Since
  $Z_0 = \sqrt{1 + 4\,l^2 / M}$, this root clearly depends upon
  $l$. We have computed this dependence and summarized the results
  in table~\ref{table2}~\footnote{It is interesting to note 
  that the roots with the
  $\ast$ in table \ref{table2} are far from precise. With only ten
  significant figures for $\nu_o$, the values of $I_{\nu_o}(Z_0)$
  for $Z_0 = 20$ and $Z_0 = 30$ are of order $O(10^2)$ and
  $O(10^{9})$ respectively. This is the way the modified Bessel
  function behaves for large $Z_0$. In the other cases, the
  corresponding values of $I_{\nu_o}(Z_0)$ are of order at most
  $O(10^{\minus 5})$. Indeed one will need more than 25
  significant figures in order to bring $I_{\nu_o}(30)$ to the
  order $O(10^{\minus 4})$.}
  , where $\nu_o \equiv \minus i\,\omega_o / \lambda$. As $l$
  increases (i.e. $Z_0$ increases), the real part of
  $\minus \nu_o$ also increases, yielding a more rapid falloff
  rate.
  \begin{table}
    \hfill \vbox{\tabskip=0pt \offinterlineskip
	\def\tablerule{\noalign{\hrule}} \halign to260pt{
		\strut #& \vrule # \tabskip=2mm plus2mm &
		\hfil # \hfil & \vrule # &
		\hfil # \hfil & \vrule # \tabskip=0pt \cr \tablerule
		&& $Z_0$ && $\nu_o \equiv \minus i\,\omega_o/\lambda$ & \cr \tablerule
		&& 1  && $\minus 1.296655185$ & \cr \tablerule
		&& 3  && $\minus 2.246635543-1.620352348\,i$ & \cr \tablerule
		&& 5  && $\minus 2.689085439-3.379067026\,i$ & \cr \tablerule
		&& 10 && $\minus 3.416051394-7.973977837\,i$ & \cr \tablerule
		&& 20 $\ast$ && $\minus 4.325727136-17.46003536\,i$ & \cr \tablerule
		&& 30 $\ast$ && $\minus 4.961702418-27.09827714\,i$ & \cr \tablerule
    }} \hfill \mbox{}
    \caption{The roots for $I_{\nu_o}(Z_0) = 0$ that are
	     responsible for the dominant exponential falloff}
    \label{table2}
  \end{table}

  This leads to a qualitative distinction between the mass
  inflation mechanism in the (static) $(2+1)$ and
  $(3+1)$-dimensional spacetimes. In a Schwarzschild background,
  the falloff rate is  $1 / t^{2\,l+3}$ \cite{Price} which implies
  larger the moment $l$, faster the falloff. This yields a mass
  inflation rate  $\sim \exp(\kappa\,v)/v^{2l+3}$ \cite{Poisson};
  the exponential growth always surpasses the falloff effect for
  arbitrarily large $l$. For a 3D black hole with an inner horizon
  our results imply a falloff rate $\sim \exp( (\kappa-\alpha(l))\,v )$.
  Therefore it is not difficult to imagine a situation in which
  $l$ is so large that the magnitude of the imaginary part of
  $\omega_o$ yields an $\alpha$ which exceeds the surface gravity
  $\kappa$ which is independent of $l$ at the Cauchy horizon,
  cutting off mass inflation for those modes.

\section*{Acknowledgements}
  This work was supported in part by the Natural Sciences and
  Engineering Research Council of Canada.

\appendix
  \section{Solutions in asymptotically flat $(D+1)$-dimensional background}
  \label{D+1apdx}
    Given the lapse function (\ref{DLapse}) in $(D+1)$-dimensional
    spacetime, the time-independent solution of $\SOL{\psi(r)}=0$
    is given by equation (\ref{DStatic}). The coefficients
    $a_j(r)$ and $c_j(r)$ can be generated by the next two
    equations:
    \begin{eqnarray}
      a_{j+1}(r) & = &
      m\,\RdBk{\ln\!|r|}^\beta\,a_j(r) \nonumber \\&&
      - \; \frac {m\,\SqBk{\xi\,(D-2-\alpha)\,(D-1-\alpha)+l\,(l+D-2-\alpha)}}{2\,\gamma+1}\,\W{1+(j+1)\,\alpha}{\RdBk{\ln\!|r|}^\beta\,a_j(r)} \nonumber \\&&
      + \; \frac {m\,\SqBk{\xi\,(D-2-\alpha)\,(D-1-\alpha)+(l+\alpha)\,(l+D-2)}}{2\,\gamma+1}\,\W{2\,\gamma+2+(j+1)\,\alpha}{\RdBk{\ln\!|r|}^\beta\,a_j(r)} \nonumber \\&&
      - \; \frac {m\,\beta\,\SqBk{\xi\,(2\,D-3-2\,\alpha)+l}}{2\,\gamma+1}\,\W{1+(j+1)\,\alpha}{\RdBk{\ln\!|r|}^{\beta-1}\,a_j(r)} \nonumber \\&&
      + \; \frac {m\,\beta\,\SqBk{\xi\,(2\,D-3-2\,\alpha)-l-D+2}}{2\,\gamma+1}\,\W{2\,\gamma+2+(j+1)\,\alpha}{\RdBk{\ln\!|r|}^{\beta-1}\,a_j(r)} \nonumber \\&&
      - \; \frac {m\,\beta\,(\beta-1)\,\xi}{2\,\gamma+1}\,\W{1+(j+1)\,\alpha}{\RdBk{\ln\!|r|}^{\beta-2}\,a_j(r)} \nonumber \\&&
      + \; \frac {m\,\beta\,(\beta-1)\,\xi}{2\,\gamma+1}\,\W{2\,\gamma+2+(j+1)\,\alpha}{\RdBk{\ln\!|r|}^{\beta-2}\,a_j(r)} \comma \\
      c_{j+1}(r) & = &
      m\,\RdBk{\ln\!|r|}^\beta\,c_j(r) \nonumber \\&&
      + \; \frac {m\,\SqBk{\xi\,(D-2-\alpha)\,(D-1-\alpha)+(l+\alpha)\,(l+D-2)}}{2\,\gamma+1}\,\W{1+(j+1)\,\alpha}{\RdBk{\ln\!|r|}^\beta\,c_j(r)} \nonumber \\&&
      - \; \frac {m\,\SqBk{\xi\,(D-2-\alpha)\,(D-1-\alpha)+l\,(l+D-2-\alpha)}}{2\,\gamma+1}\,\W{\minus 2\,\gamma+(j+1)\,\alpha}{\RdBk{\ln\!|r|}^\beta\,c_j(r)} \nonumber \\&&
      + \; \frac {m\,\beta\,\SqBk{\xi\,(2\,D-3-2\,\alpha)-l-D+2}}{2\,\gamma+1}\,\W{1+(j+1)\,\alpha}{\RdBk{\ln\!|r|}^{\beta-1}\,c_j(r)} \nonumber \\&&
      - \; \frac {m\,\beta\,\SqBk{\xi\,(2\,D-3-2\,\alpha)+l}}{2\,\gamma+1}\,\W{\minus 2\,\gamma+(j+1)\,\alpha}{\RdBk{\ln\!|r|}^{\beta-1}\,c_j(r)} \nonumber \\&&
      + \; \frac {m\,\beta\,(\beta-1)\,\xi}{2\,\gamma+1}\,\W{1+(j+1)\,\alpha}{\RdBk{\ln\!|r|}^{\beta-2}\,c_j(r)} \nonumber \\&&
      - \; \frac {m\,\beta\,(\beta-1)\,\xi}{2\,\gamma+1}\,\W{\minus 2\,\gamma+(j+1)\,\alpha}{\RdBk{\ln\!|r|}^{\beta-2}\,c_j(r)} \period
    \end{eqnarray}
    The function $W$ used above is defined as
    \begin{eqnarray*}
      \W{n}{f(r)} & \equiv & \frac{r^n}{r}\,\int \frac{f(r)}{r^n}\,dr \comma
    \end{eqnarray*}
    where $n$ is a real number and $f(r)$ is an integrable
    function. We evaluate the integral in $W$ in such a way that
    the integration constant is always set to zero. This function
    has a property that if $f(r)$ is a constant and $n \neq 1$,
    then $\W{n}{f(r)}$ gives a constant. When $n$ is unity but
    $f(r)$ is still a constant, $\W{n}{f(r)}$ is proportional to
    $\ln\!|r/{\cal R}|$, where ${\cal R}$ is some constant of
    dimension of length. In the case when $f(r)$ is no longer
    constant but a polynomial of $\ln\!|r/{\cal R}|$, the function
    $W$ gives another polynomial of $\ln\!|r/{\cal R}|$. The index
    $j$ in the two equations runs from zero to infinity and the
    starting coefficients $a_0(r)$ and $c_0(r)$ are two arbitrary
    constants. It is obvious that when the background is flat,
    i.e. $m = 0$, all the $a_j(r)$ and $c_j(r)$ vanish except
    $a_0$ and $c_0$. If $m \neq 0$, these coefficients are
    polynomials in $\ln\!|r/{\cal R}|$.

    For the equation $\SOL{C_{i+1}(r)} = 2\,d\,C_i(r)/d\,r$ with
    $\SOL{C_0(r)} = 0$, the solution of $C_i(r)$ is given by
    \begin{eqnarray*}
      C_i(r) & = &
      r^{\gamma+1+i}\,\sum_{j=0}^\infty \frac{c^i_j(r)}{r^{j\,\alpha}} \period
    \end{eqnarray*}
    Each coefficient $c^i_j(r)$ can be calculated by the next two
    equations.
    \begin{eqnarray*}
      c^i_0(r) & = &
      c^i_0 \; = \;
      \frac{2^i\,(2\,\gamma+1)!\,(\gamma+i)!}{\gamma!\,(\,i\,)!\,(2\,\gamma+1+i)!}\,c_0 \comma \\
      c^{i+1}_{j+1}(r) & = &
      m\,\RdBk{\ln\!|r|}^\beta\,c^{i+1}_j(r) \nonumber \\&&
      + \; \frac{2}{2\,\gamma+1}\,\SqBk{\gamma\,\W{\minus i+(j+1)\,\alpha}{c^i_{j+1}(r)}+(\gamma+1)\,\W{\minus 2\,\gamma-1-i+(j+1)\,\alpha}{c^i_{j+1}(r)}} \nonumber \\&&
      + \; \frac {m\,\SqBk{\xi\,(D-2-\alpha)\,(D-1-\alpha)+(l+\alpha)\,(l+D-2)}}{2\,\gamma+1}\,\W{\minus i+(j+1)\,\alpha}{\RdBk{\ln\!|r|}^\beta\,c^{i+1}_j(r)} \nonumber \\&&
      - \; \frac {m\,\SqBk{\xi\,(D-2-\alpha)\,(D-1-\alpha)+l\,(l+D-2-\alpha)}}{2\,\gamma+1}\,\W{\minus 2\,\gamma-1-i+(j+1)\,\alpha}{\RdBk{\ln\!|r|}^\beta\,c^{i+1}_j(r)} \nonumber \\&&
      + \; \frac {m\,\beta\,\SqBk{\xi\,(2\,D-3-2\,\alpha)-l-D+2}}{2\,\gamma+1}\,\W{\minus i+(j+1)\,\alpha}{\RdBk{\ln\!|r|}^{\beta-1}\,c^{i+1}_j(r)} \nonumber \\&&
      - \; \frac {m\,\beta\,\SqBk{\xi\,(2\,D-3-2\,\alpha)+l}}{2\,\gamma+1}\,\W{\minus 2\,\gamma-1-i+(j+1)\,\alpha}{\RdBk{\ln\!|r|}^{\beta-1}\,c^{i+1}_j(r)} \nonumber \\&&
      + \; \frac {m\,\beta\,(\beta-1)\,\xi}{2\,\gamma+1}\,\W{\minus i+(j+1)\,\alpha}{\RdBk{\ln\!|r|}^{\beta-2}\,c^{i+1}_j(r)} \nonumber \\&&
      - \; \frac {m\,\beta\,(\beta-1)\,\xi}{2\,\gamma+1}\,\W{\minus 2\,\gamma-1-i+(j+1)\,\alpha}{\RdBk{\ln\!|r|}^{\beta-2}\,c^{i+1}_j(r)} \period
    \end{eqnarray*}

\end{document}